\documentclass[10pt]{llncs}

\usepackage{lmodern}
\usepackage[english]{babel}
\usepackage{booktabs}
\usepackage{url,xspace}
\usepackage{multirow}

\usepackage{tikz}
\usetikzlibrary{calc, positioning, arrows, backgrounds, matrix}
\usetikzlibrary{shapes}

\tikzset{
  DOp/.style={
    draw, rectangle
  },
  LOp/.style={
    draw, rounded rectangle
  },
  Action/.style={
    draw, chamfered rectangle, inner sep=2pt
  },
}

\usepackage[T1]{fontenc}
\usepackage[utf8]{inputenc}
\usepackage{csquotes}
\usepackage{textcomp}
\usepackage{amssymb}
\usepackage{amsmath}

\usepackage{hyperref}

\hypersetup{
  colorlinks=false,
  hidelinks,
  pdfborder={0 0 0},
  bookmarksopen=true,
  bookmarksnumbered=true,
  breaklinks=true,
}

\usepackage[paper=17cm:24cm,DIV=16,BCOR=10.8mm,headinclude,pagesize]{typearea}

\usepackage{bbm}

\makeatletter
\DeclareRobustCommand*{\bfseries}{%
  \not@math@alphabet\bfseries\mathbf
  \fontseries\bfdefault\selectfont
  \boldmath\let\mathbbnormal=\mathbb\let\mathbb=\mathbbm
}
\makeatother

\usepackage[algo2e,ruled,vlined,linesnumbered]{algorithm2e}
\DontPrintSemicolon
\SetAlCapHSkip{0pt}
\IncMargin{-\parindent}
\SetCommentSty{textit}
\SetProgSty{}
\SetFuncSty{}
\SetArgSty{}

\SetKwProg{Function}{function}{}{}

\SetKwFunction{CmpName}{CmpName}
\SetKwFunction{Filter}{Filter}
\SetKwFunction{FlatWindow}{FlatWindow}
\SetKwFunction{Map}{Map}
\SetKwFunction{Max}{Max}
\SetKwFunction{Merge}{Merge}
\SetKwFunction{PrefixSum}{PrefixSum}
\SetKwFunction{Size}{Size}
\SetKwFunction{Sort}{Sort}
\SetKwFunction{Union}{Union}
\SetKwFunction{Window}{Window}
\SetKwFunction{ZipWithIndex}{ZipWithIndex}
\SetKwFunction{Zip}{Zip}
\SetKwFunction{Unique}{Unique}
\SetKwFunction{Discard}{NPairs}
\SetKwFunction{Cache}{Cache}
\SetKwFunction{Collapse}{Collapse}

\def\KwIfThenElse#1#2#3{%
  (\KwSty{if }{#1}\KwSty{ then }{#2}\KwSty{ else }{#3})%
}

\newcommand{\Rem}[1]{\tcp*{#1}}
\newcommand{\Remi}[1]{\tcp*[f]{#1}}

\newcommand{\arr}[1]{[\, #1 \,]}

\usepackage{microtype}

\newcommand{\UnaryMathOperator}[4][]{%
  \if\relax\detokenize{#1}\relax%
  \ensuremath{\mathop{}\mathopen{}#2\mathopen{}}%
  \else%
  \ensuremath{\mathop{}\mathopen{}#2\mathopen{}#3#1#4}%
  \fi%
}

\newcommand{\UnaryOperator}[4][]{%
  \if\relax\detokenize{#1}\relax%
  \ensuremath{\mathop{}\mathopen{}#2\mathopen{}}%
  \else%
  \ensuremath{\mathop{}\mathopen{}#2\mathopen{}#3}#1\ensuremath{#4}%
  \fi%
}

\newcommand{\DIA}[1]{\UnaryMathOperator[#1]{\texttt{DIA}}{\langle}{\rangle}}

\def\SA{\textsf{SA}}
\def\ISA{\textsf{ISA}}

\begin{document}

\title{Scalable Construction of Text Indexes}

\author{
Timo Bingmann\inst{1}\and
Simon Gog\inst{1}\and
Florian Kurpicz\inst{2}
}

\institute{
Institute of Theoretical Informatics,\\
Karlsruhe Institute of Technology, Germany
\and 
Department of Computer Science,\\
Technische Universit\"at Dortmund, Germany
}

\maketitle


\begin{abstract}
The suffix array is the key to efficient solutions for myriads of string processing problems in different applications domains, like data compression, data mining, or Bioinformatics.
With the rapid growth of available data, suffix array construction algorithms had to be adapted to advanced computational models such as external memory and distributed computing.
In this article, we present five suffix array construction algorithms utilizing the new algorithmic big data batch processing framework Thrill, which allows us to process input sizes in
orders of magnitude that have not been considered before.
\end{abstract}


\section{Introduction}

Suffix arrays~\cite{manber1993suffix,gonnet1992new} are the basis for many text indexes and string algorithms.
Suffix array construction is theoretically linear work, but practical suffix sorting is computationally intensive and often limits the applicability of advanced text data structures on large datasets.
While fast sequential algorithms exist in the RAM model~\cite{mori2006divsufsort,nong2009linear}, these are limited by the CPU power and RAM size of a single machine.
External memory algorithms on a single machine are limited by disk~\cite{dementiev08better,bingmann2013inducing,karkkainen2015parallel}, and often have long running times due to mostly sequential computation or limited I/O bandwidth.

Most suffix array construction algorithms focus only on sequential computation models.
However, while the volume of data is increasing, the speed of individual CPU cores is not.
This leaves us no choice but to consider shared-memory parallelism and distributed cluster computation to gain considerable speedups in the future.

Most suffix array construction algorithms (SACAs) employ a subset of three basic suffix sorting principles: \emph{prefix doubling}, \emph{recursion} and \emph{inducing}~\cite{puglisi07taxonomy}.
The last type, \emph{inducing}, is the basis for the fastest \emph{sequential} suffix array construction algorithms \cite{mori2006divsufsort,nong2009linear}, but yields only well to parallelization for small alphabets~\cite{labeit2015parallel}, and does not appear to be a promising approach for distributed environments.
Recently, a fast distributed prefix doubling implementation using MPI has been presented~\cite{flick2015parallel}.
While they report high speeds for very small inputs, we could not successfully run their implementation on large inputs.
Furthermore, using hundreds of high performance machines for small inputs is not dollar-cost-efficient.

We propose to use the big data framework \emph{Thrill}, which supports distributed external memory algorithms for suffix sorting of large inputs.

After giving a short introduction to Thrill in Section~\ref{ssec:a_short_introduction_into_thrill}, we provide a detailed description of our SACA implementations in Section~\ref{sec:scalabel_suffix_array_construction_algorithms}. Section~\ref{ssec:prefix_doubling_algorithms} considers multiple variants of prefix doubling algorithms, and Section~\ref{ssec:difference_cover_algorithms} discusses the recursive difference cover algorithms DC3 and DC7.


\subsection{Related Work}
There exists numerous work on sequential SACAs, see \cite{puglisi07taxonomy,Dhaliwal2012} for two overview articles.
Research in this area is still active, as just this year another theoretically optimal SACA has been presented that combines ideas used in prefix doubling and inducing~\cite{Baier2016}.
K{\"a}rkk{\"a}inen et al.~\cite{karkkainen2006linear,karkkainen2003simple} presented a linear time SACA, the so called DC3 algorithm, that works well in multiple advanced models of computation such as external memory and also parallel and distributed environments.
Kulla and Sanders showed the scalability of the DC3 algorithm in a distributed environment~\cite{kulla2007scalable}.
More recently, Flick and Aluru presented an implementation of a prefix doubling algorithm in MPI that can also compute the longest common prefix array~\cite{flick2015parallel}.
SACAs have also been considered in external memory, where in theory the DC3 algorithm~\cite{karkkainen2006linear} is optimal.
Dementiev et al.~\cite{dementiev08better} compared multiple implementations of prefix doubling and DC3 for external memory in practice.
Lately, K{\"a}rkk{\"a}inen et al.~\cite{karkkainen2015parallel,Karkkainen2014} presented two differente external memory SACAs.

Related to SACAs are construction algorithms for the suffix tree and the Burrows-Wheeler transform (BWT), which are often used in Bioinformatics.
In this domain one can, however, make special assumptions such as that the input text is fairly random (like DNA), or that one wishes to compress multiple very similar texts (like multiple genome sequences of the same species).
These practical assumptions yield suffix sorting implementations tailored to their applications, like straight-forward parallel radix sort~\cite{mansour2011era,wang2015bwtcp} or merging of multiple BWTs generated in parallel~\cite{siren16dcc}.
On general text these implementations, however, have super-linear theoretical running time.


\subsection{A Short Introduction into Thrill}
\label{ssec:a_short_introduction_into_thrill}
We implemented five suffix array construction algorithms using the distributed big data batch computation framework \emph{Thrill}~\cite{bingmann2016thrill}.
Thrill works with \emph{distributed immutable arrays} (\DIA{}s) storing tuples.
Items in \DIA{}s cannot be accessed directly, instead Thrill provides a rich set of \DIA{} operations which can be used to transform \DIA{}s (we use and describe only a subset of the operations Thrill provides).
Each \DIA{} operation can be instantiated with appropriate user-defined functions for constructing complex algorithms.

\begin{description}
\item[$\operatorname{Filter}(f)$]
  takes a \DIA{A} $X$ and a function $f \colon A \to \texttt{bool}$, and returns the \DIA{A} containing $\arr{ x \in X \mid f(x) }$ within which the order of items is maintained.

\item[$\operatorname{Map}(f)$]
  applies the function $f \colon A \to B$ to each item in the input \DIA{A} $X$, and returns a \DIA{B} $Y$ with $Y[i]=f(X[i])$ for all $i = 0,\dots, |X| - 1$.

\item[$\operatorname{Window}_k(w)$ and $\operatorname{FlatWindow}_k(w')$]
  takes an input \DIA{A} $X$ and a window function $w \colon \mathbb{N}_0 \times A^k \to B$.
  The operation scans over $X$ with a window of size $k$ and applies $w$ once to each set of $k$ consecutive items from $X$ and their index in $X$.
  The final $k - 1$ indexes with less than $k$ consecutive items are delivered to $w$ as partial windows padded with sentinel values.
  The result of all invocations of $w$ is returned as a \DIA{B} containing $|X|$ items in the order.

  $\FlatWindow$ is a variant of $\Window$ which takes a input \DIA{A} $X$ and a window function $w' \colon \mathbb{N}_0 \times A^k \to \text{list}(B)$.
  The only difference compared to Window is, that $w'$ can \emph{emit} zero or more items that are concatenated in the resulting $\DIA{B}$ in the order they are emitted.\footnote{We say the items are \emph{emitted}, as in other \DIA{} operations more than one item can be created per call of the function $w$ to the output \DIA{B}, while \emph{return} exits the function.}

\item[$\operatorname{PrefixSum}(s)$]
  Given an input \DIA{A} $X$ and an associative operation $s \colon A \times A \to A$ (by default $s = +$), PrefixSum returns a \DIA{A} $Y$ such that $Y[0] = X[0]$ and $Y[i] = s(Y[i-1], X[i])$ for all $i = 1, \dots, |X| - 1$.

\item[$\operatorname{Sort}(c)$]
  sorts an input \DIA{A} $X$ with respect to a less-comparison function $c \colon A \times A \to \text{bool}$.
  If Sort is called without a comparison function, we assume the tuples are compared component-wise with the first component being most significant, the second component the second most significant, and so on.

\item[$\operatorname{Merge}(X_1,\dots,X_n,c)$]
  Given a set of sorted \DIA{A}s $X_1,\dots,X_n$ and a less-com\-parison function $c \colon A \times A \to \text{bool}$, Merge
  returns \DIA{A} $Y$ that contains all tuples of $X_1,\dots,X_n$ and is sorted with respect to $c$.
  If Merge is called without a comparison function we compare the tuples component-wise (see Sort).

\item[$\operatorname{Union}(X_1,\dots,X_n)$]
  Given a set of \DIA{A}s $X_1,\dots,X_n$, Union returns \DIA{A} $Y=\bigcup_{i=1}^{n} X_i$ containing all items of the input in an arbitrary order.

\item[$\operatorname{Zip}(X_1,\dots,X_n,f)$]
  Given a set of \DIA{}s $X_1,\dots,X_n$ of type $A_1,\dots,A_n$ of equal size ($|X_1| = \dots = |X_n|$) and a function $f\colon A_1 \times \dots \times A_n \to B$,
  Zip returns \DIA{B} $Y$ with $Y[i] = f(X_1[i],\dots,X_n[i])$ for all $i = 0, \dots, |X_1| - 1$.

\item[$\operatorname{ZipWithIndex}(f)$]
  Given an input \DIA{A} $X$ and a function $f : (\mathbb{N}_0,A) \rightarrow B$, ZipWithIndex returns \DIA{B} $Y$ with $Y[i] = f(i,X[i])$ for all $i=0,\dots,|X| - 1$

\item[$\operatorname{Max}(c)$]
  Given an input \DIA{A} $X$, Max returns the maximum item $m=\max_c X$ with respect to a less-comparison function $c \colon A \times A \to \text{bool}$. By default (if Max is called without a comparison function) the tuples are compared component-wise (see Sort).

\item[$\operatorname{Size}()$]
  Given an input \DIA{A} $X$, Size returns the number of items in $X$, i.e., $|X|$.

\end{description}

Thrill applies chains of functions (\emph{method chaining}) to a \DIA{}, e.g., if we have a $\DIA{\mathbb{N}_0}~N=\lbrace 0,1,2,\dots,9  \rbrace$ and want to compute the prefix sum of all odd elements, then we write $N.\Filter(a \mapsto (a \mod 2 = 1).\PrefixSum()$.
Using chaining, the operations form a data-flow style graph of \DIA{} operations.
Drawings of this graph help to give a visual impression of the dependencies between the operations.
As the data-flow drawings in this paper are generated from our actual Thrill implementation, they contain some additional nodes.
These are only needed for performance (\textbf{Cache}) and due to the way Thrill code is chained (\textbf{Collapse}).


\section{Scalable Suffix Array Construction Algorithms}
\label{sec:scalabel_suffix_array_construction_algorithms}
In this section, we describe the suffix array construction algorithms that we have implemented in Thrill.
First we describe algorithms based on prefix doubling, and then we present two implementations based on recursion.
SACAs based on inducing do not appear to be a promising approach in a distributed environment.

Given is a text $T$ of length $|T| = n$ over an alphabet $\Sigma$.
We call the substring $T[i,n)$ the $i$-th \emph{suffix} of $T$.
The \emph{suffix array} (\SA) for $T$ is a permutation of $[0, n)$ such that $T[\SA[i],n) \leq_{\text{lex}} T[\SA[j],n)$ for all $0 \leq i \leq j < n$.
The inverse permutation of SA is called the \emph{inverse suffix array} (\ISA) and the \emph{lexicographic rank} of the $i$-th suffix is $\ISA[i]$.
While the ranks of all suffixes are distinct, we will often use the notion of a \emph{lexicographic name}.
Lexicographic names are representatives of suffixes which need not be distinct but do respect the lexicographic ordering, i.e., $n_i$ and $n_j$ are lexicographic names of two suffixes $T[i,n) < T[j,n)$ iff $n_i \leq n_j$.

\subsection{Prefix Doubling Algorithms}
\label{ssec:prefix_doubling_algorithms}
The goal of a \emph{prefix doubling} algorithm is to give each suffix of $T$ a lexicographic name such that the name corresponds to the rank of the suffix in the (partial) \SA.
The names are computed using prefixes of length $2^{k}$ of the suffixes for $k=1,\dots,\lceil \log_2 |T|\rceil$.
During each step, we double the length of these prefixes (hence the name of this type of algorithm).
We can compute the name for the prefix $T[i, i+2^{k})$ using the already computed names of the prefixes $T[i,i+2^{k-1})$ and $T[i+2^{k-1},i+2^{k})$.

\begin{algorithm2e}[b]
  \caption{Generic Prefix Doubling algorithm.\label{alg:prefix_doubling_generic}}
  \Function{PrefixDoubling($T \in \DIA{\Sigma}$)}{
    $S:= T.\Window_2((i, \arr{t_0,t_1}) \mapsto (i,t_0,t_1))$
    \Rem{Create initial triples $(i,T[i], T[i+1])$.}\label{alg:generic_initial_tuples}
    \For{$k:= 1$ \KwSty{to} $\lceil \log_2 |T| \rceil - 1$}{\label{alg:generic_loop}
      $S:= S.\Sort((i,r_0,r_1) \text{ by } (r_0,r_1))$ \Rem{Sort triples by name pair.}\label{alg:generic_sort_name_pair}
      $N:= S.\FlatWindow_2((i, \arr{a, b}) \mapsto \CmpName{i, a, b})$ \Rem{Map to names $0$ or $i$.}\label{alg:generic_map_name}
      \If(\Remi{If all names distinct, then}){$N.\Filter((i, r) \mapsto (r = 0)).\Size() = 1$}{\label{alg:generic_finished}
        \Return $N.\Map((i, r) \mapsto i)$ \Rem{return names as suffix array,}\label{alg:generic_return}
      }
      $N:= N.\PrefixSum((i,r), (i^\prime,r^\prime) \mapsto (i^\prime, \max(r, r^\prime))$
      \Rem{else calculate new names}\label{alg:generic_new_name}
      $S:= \textbf{Generate new name pairs using N}$
      \Rem{and run next refinement iteration.}\label{alg:generic_new_tuples}
    }
  }
\end{algorithm2e}

\begin{figure}[t]
  \centering
  \begin{tikzpicture}[scale=0.4]
    \node (1) at (90bp,1098bp) [] {$T$};
    \node (2) at (90bp,1026bp) [DOp] {$S := T.\Window_2$};
    \node (3) at (90bp,954bp) [DOp] {$S := S.\Sort$};
    \node (4) at (90bp,882bp) [DOp] {$N := S.\FlatWindow_2$};
    \node (5) at (0bp,810bp) [LOp] {$N.\Filter$};
    \node (6) at (0bp,738bp) [Action] {$\Size$};
    \node (7) at (180bp,810bp) [DOp] {$N := N.\PrefixSum$};
    \node (8) at (180bp,738bp) [DOp] {$N := N.\Sort$};
    \node (9) at (90bp,666bp) [DOp] {$S := N.\Window_2$};
    \node (10) at (90bp,594bp) [DOp] {$S := S.\Sort$};
    \node (11) at (90bp,522bp) [DOp] {$N := S.\FlatWindow_2$};
    \node (12) at (0bp,450bp) [LOp] {$N.\Filter$};
    \node (13) at (0bp,378bp) [Action] {$\Size$};
    \node (14) at (192bp,450bp) [LOp] {$N.\Map$};
    \node (15) at (192bp,378bp) [] {$\SA_T$};
    \draw [->] (1) -- (2);
    \draw [->] (2) -- (3);
    \draw [->] (3) -- (4);
    \draw [->] (4) -- (5);
    \draw [->] (4) -- (7);
    \draw [->] (5) -- (6);
    \draw [->] (7) -- (8);
    \draw [->] (8) -- (9);
    \draw [->] (9) -- (10);
    \draw [->] (10) -- (11);
    \draw [->] (11) -- (12);
    \draw [->] (11) -- (14);
    \draw [->] (12) -- (13);
    \draw [->] (14) -- (15);
  \end{tikzpicture}
  \caption{DIA data-flow graph of two iterations of prefix doubling with sorting.}\label{fig:dataflow-pd}
\end{figure}
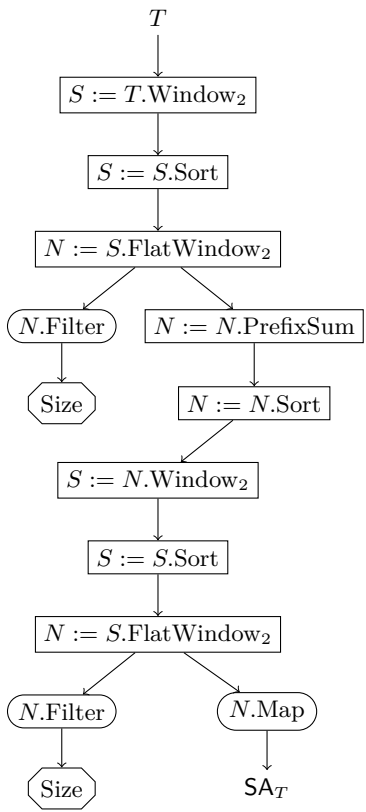

Algorithm~\ref{alg:prefix_doubling_generic} describes the basic structure of the prefix doubling algorithms presented in this section.
The corresponding data-flow graph is shown in Figure~\ref{fig:dataflow-pd}.
The whole algorithm requires one \DIA{} $N$ storing tuples and one \DIA{} $S$ storing triples.
Initially,  $S$ contains the triples $(i, T[i], T[i + 1])$ for all $i = 0,\dots, n - 1$ where we assume that $T[n] = \texttt{\$}$, see Line~\ref{alg:generic_initial_tuples}.
These triples contain a text position and the \emph{name pair} for that position, i.e, the two names that are required to compute the new name for the suffix starting at the text position.
Next in Line~\ref{alg:generic_sort_name_pair}, we sort $S$ with respect to the name pair as we know that the names correspond to the ranks of the suffixes.
Now we prepare the computation of the new names using the functions \CmpName{} that takes the current position $i$ in $S$ and the items $S[i]$ and $S[i + 1]$ as input and emits a tuple consisting of a text position and a new name, see Algorithm~\ref{alg:prefix_doubling_name}.
We know that the suffixes are sorted with respect to their name pairs. Therefore, we can scan $S$ and mark every position where the name pair differs from its predecessor.
\CmpName{} marks these non-unique names pairs by giving them the name $0$.
All unique names pairs get a name equal to their current position in $S$.
If there is only one suffix with name $0$ we know that all names differ and that we have finished the computation, see Line~\ref{alg:generic_finished}.
Otherwise, we can use a the \DIA{} operation \PrefixSum{} to set the name of the tuple to the largest preceding name, i.e., the the new name which is unique if the name was not $0$ and the preceeding name is not $0$, see Line~\ref{alg:generic_new_name}.
Now each suffix has a new, more refined name.
The next step (see Line~\ref{alg:generic_new_tuples}) is to identify the ranks of the suffixes required for the next doubling step.
During the $k$-th doubling step, we fill $S$ with one triple for each index $i=0,\dots,|T|-1$ that contains the current name of the suffix at position $i$ and the current name of the suffix at position $i+2^{k-1}$.
This is also the step, where the prefix doubling algorithms presented here differ.
Next, we show two different approaches to compute the name pairs for the next refining iteration.

\begin{algorithm2e}[t]
  \caption{Identifications of suffix array intervals.}\label{alg:prefix_doubling_name}
  \Function{\CmpName{$j \in \mathbb{N}_0$, $(i,r_0,r_1), (i^\prime, r_0^\prime, r_1^\prime) \in N$}}{
    \If{$j = 0$}{
      $\KwSty{emit}~(i,0)$ \Rem{First \DIA{} item has no offset.}
    }
    $\KwSty{emit}~\begin{cases}
      (i^\prime, j) & \text{if } (r_0,r_1) \neq (r_0^\prime,r_1^\prime),\Rem{Add sentinel if rank pairs alter.} \\
      (i^\prime, 0) & \text{otherwise.} \Rem{$T[i,n)$ and $T[i^\prime,n)$ get the same new name.}
    \end{cases}$
  }
\end{algorithm2e}

\subsubsection{Prefix Doubling using Sorting.}

In the seminal suffix array paper by Manber and Myers~\cite{manber1993suffix}, the presented SACA was a prefix doubling algorithm using sorting.
This idea was refined by Dementiev et al.~\cite{dementiev08better} who presented an external memory SACA that we adapted to Thrill.
The idea is to compute the new name pairs by sorting the old names with respect to the starting position of the suffix, see Algorithm~\ref{alg:prefix_doubling_sort}.
We make use of the fact that during each iteration we know for each suffix the suffix whose current name is required to compute the new, refined name.
Hence, we can sort the tuples containing the starting positions of the suffixes and their current name in such a way that if there is another name required for a name pair, then it is the name of the succeeding tuple, see Line~\ref{alg:sorting_sort}.
To do so, we use the following less-comparator $<^{k}_{\mathrm{op}}\colon (\mathbb{N}_0, \mathbb{N}_0)\times (\mathbb{N}_0, \mathbb{N}_0) \to \mathtt{bool}$ (see Equation~\ref{eq:less_comparator}) in Algorithm~\ref{alg:prefix_doubling_sort}:

\begin{equation}\label{eq:less_comparator}
(i, r)<^{k}_{\mathrm{op}} (i^\prime, r^\prime) =
\begin{cases}
  i \textbf{ div } 2^k < i^\prime \textbf{ div } 2^k & \text{if } i \equiv i^\prime \pmod{2^k} \,,\\
  i \textbf{ mod } 2^k < i^\prime \textbf{ mod } 2^k & \text{otherwise}. \end{cases}
\end{equation}

After sorting using the $<^{k}_{\mathrm{op}}$-comparator, we need to ensure that two consecutive names are the ones required to compute the new name, since the required name may not exist due to the length of the text.
This occurs during the $k$-th iteration for each suffix beginning at a text position greater than $n - 2^k$.
In this case we use the sentinel name $0$ which compares smaller than any valid name, see Line~\ref{alg:sorting_exist_next}.
In both cases, we return one triple for each position, consisting of a text position, the current name of the suffix beginning at that position and the name of the suffix $2^{k}$ positions to the right (if it exists and $0$ otherwise).

\begin{algorithm2e}[t]
  \caption{Prefix Doubling using sorting.\label{alg:prefix_doubling_sort}}
  \Function{PrefixDoublingSorting($k \in \mathbb{N}_0$)}{
    $N := N.\Sort(<^{k}_{\mathrm{op}})$
    \Rem{Sort such that names required for renaming are consecutive.}\label{alg:sorting_sort}
    $S := N.\Window_2\left((j, \arr{(i,r_0,r_1), (i^\prime,r_0^\prime,r_1^\prime)}) \mapsto \begin{cases}
        (i, r_0, r_0^\prime) & \text{if } i + 2^{k} = i^\prime \,,\\
        (i, r_0, 0)     & \mathrm{otherwise}.\end{cases}
    \right)$\label{alg:sorting_exist_next}
  }
\end{algorithm2e}

Now we give an example of prefix doubling using sorting in Thrill, see Example~\ref{alg:generic_prefix_doubling_sorting_example}.
We compute the suffix array of the text $T=\texttt{bdacbdacb}$.
The comment at the end of each line refers to the line of code responsible for the change from the previous line where $x.y$ denotes line $y$ in Algorithm $x$.

\SetAlgorithmName{Example}{example}{List of Examples}
\begin{algorithm2e}[t]
  \def\C#1{\texttt{#1}}
  \caption{Example of prefix doubling using sorting in Thrill.\label{alg:generic_prefix_doubling_sorting_example}}
  $T = \arr{\C{b},\C{d},\C{a},\C{c},\C{b},\C{d},\C{a},\C{c},\C{b} }$\;
  $S = \arr{ (0,\C{d},\C{b}), (1,\C{b},\C{a}), (2,\C{a},\C{c}), (3,\C{c},\C{b}), (4,\C{b},\C{d}), (5,\C{d},\C{a}),
             (6,\C{a},\C{c}), (7,\C{c},\C{b}), (8,\C{b},\C{\$}) }$
             \Rem{\ref{alg:prefix_doubling_generic}.\ref{alg:generic_initial_tuples}}
  $\mathbf{k = 1}$\Rem{\ref{alg:prefix_doubling_generic}.\ref{alg:generic_loop}}
  $S = \arr{ (2,\C{a},\C{c}), (6,\C{a},\C{c}), (8,\C{b},\C{\$}), (0,\C{b},\C{d}), (4,\C{b},\C{d}), (3,\C{c},\C{b}),
             (7,\C{c},\C{b}), (1,\C{d},\C{a}), (5,\C{d},\C{a}) }$
             \Rem{\ref{alg:prefix_doubling_generic}.\ref{alg:generic_sort_name_pair}}
  $N = \arr{ (2, 0), (6, 0), (8, 2), (0, 3), (4, 0), (3, 5), (7, 0), (1, 7), (5, 0) }$
  \Rem{\ref{alg:prefix_doubling_generic}.\ref{alg:generic_map_name}}
  \textbf{4 items with rank 0}\Rem{\ref{alg:prefix_doubling_generic}.\ref{alg:generic_finished}}
  $N = \arr{ (2, 0), (6, 0), (8, 2), (0, 3), (4, 3), (3, 5), (7, 5), (1, 7), (5, 7) }$
  \Rem{\ref{alg:prefix_doubling_generic}.\ref{alg:generic_new_name}}
  $N = \arr{ (0, 3), (2, 0), (4, 3), (6, 0), (8, 2), (1, 7), (3, 5), (5, 7), (7, 5) }$
  \Rem{\ref{alg:prefix_doubling_sort}.\ref{alg:sorting_sort}}
  $S = \arr{ (0, 3, 0), (2, 0, 3), (4, 3, 0), (6, 0, 2), (8, 2, 0), (1, 7, 5), (3, 5, 7), (5, 7, 5), (7, 5, 0) }$
  \Rem{\ref{alg:prefix_doubling_sort}.\ref{alg:sorting_exist_next}}
  $\mathbf{k = 2}$\Rem{\ref{alg:prefix_doubling_generic}.\ref{alg:generic_loop}}
  $S = \arr{ (6, 0, 2), (2, 0, 3), (8, 2, 0), (0, 3, 0), (4, 3, 0), (7, 5, 0), (3, 5, 7), (1, 7, 5), (5, 7, 5) }$
  \Rem{\ref{alg:prefix_doubling_generic}.\ref{alg:generic_sort_name_pair}}
  $N = \arr{ (6, 0), (2, 1), (8, 2), (0, 3), (4, 0), (7, 5), (3, 6), (1, 7), (5, 0) }$
  \Rem{\ref{alg:prefix_doubling_generic}.\ref{alg:generic_map_name}}
  \textbf{2 items with rank 0}\Rem{\ref{alg:prefix_doubling_generic}.\ref{alg:generic_finished}}
  $N = \arr{ (6, 0), (2, 1), (8, 2), (0, 3), (4, 3), (7, 5), (3, 6), (1, 7), (5, 7) }$
  \Rem{\ref{alg:prefix_doubling_generic}.\ref{alg:generic_new_name}}
  $N = \arr{ (0, 3), (4, 3), (8, 2), (1, 7), (5, 7), (2, 1), (6, 0), (3, 6), (7, 5) }$
  \Rem{\ref{alg:prefix_doubling_sort}.\ref{alg:sorting_sort}}
  $S = \arr{ (0, 3, 3), (4, 3, 2), (8, 2, 0), (1, 7, 7), (5, 7, 0), (2, 1, 0), (6, 0, 0), (3, 6, 5), (7, 5, 0) }$
  \Rem{\ref{alg:prefix_doubling_sort}.\ref{alg:sorting_exist_next}}
  $\mathbf{k = 3}$\Rem{\ref{alg:prefix_doubling_generic}.\ref{alg:generic_loop}}
  $S = \arr{ (6, 0, 0), (2, 1, 0), (8, 2, 0), (4, 3, 2), (0, 3, 3), (7, 5, 0), (3, 6, 5),  (5, 7, 0), (1, 7, 7) }$
  \Rem{\ref{alg:prefix_doubling_generic}.\ref{alg:generic_sort_name_pair}}
  $N = \arr{ (6, 0), (2, 1), (8, 2), (4, 3), (0, 4), (7, 5), (3, 6), (5, 7), (1, 8) }$
  \Rem{\ref{alg:prefix_doubling_generic}.\ref{alg:generic_map_name}}
  \textbf{1 item with rank 0}\Rem{\ref{alg:prefix_doubling_generic}.\ref{alg:generic_finished}}
  \textbf{Result:} $\arr{ 6, 2, 8, 4, 0, 7, 3, 5, 1 }$\Rem{\ref{alg:prefix_doubling_generic}.\ref{alg:generic_return}}
\end{algorithm2e}
\SetAlgorithmName{Algorithm}{algorithm}{List of Algorithms}

\subsubsection{Prefix Doubling using the Inverse Suffix Array.}

During the $k$-th doubling step, we compute a name for each suffix and hence for each position of the text.
Algorithm~\ref{alg:prefix_doubling_isa} describes how we obtain the rank of the required suffixes using the inverse suffix array.
This approach has been considered in a distributed environment~\cite{flick2015parallel} and is based on the work of Larsson and Sadakane~\cite{larsson99faster} who proposed to use the inverse suffix for prefix doubling.
If we sort the names based on their position in the text, we get the \emph{partial} inverse suffix array (partial, as the inverse suffix array does not necessarily contain the final position of all suffixes in the SA).
Using this partial inverse suffix array, we can get the current rank of each suffix by its text position.
For each position $i$, we need the rank of the $(i+2^{k})$-th suffix. To get this rank, we scan over the \DIA{} with a window of width $2^{k}$, i.e., the same as shifting the partial inverse suffix array by $2^k$ positions and appending $0$s until its length is $|T|$ again.

Again, we give an example of the algorithm for the same text as before, see Example~\ref{alg:generic_prefix_doubling_isa_example}.
Also, the comments refer to the algorithm and line responsible for the change as in the previous example.

\begin{algorithm2e}[t]
  \caption{Prefix Doubling using the inverse suffix array.\label{alg:prefix_doubling_isa}}
  \Function{PrefixDoublingISA($k \in \mathbb{N}_0$)}{
    $N := N.\Sort(((i,r) \text{ by } i)$\Rem{Compute partial ISA.}\label{alg:isa_sort}
    $S := N.\Window_{2^{k}+1}\left((j, \arr{(i,r),\dots,(i^\prime,r^\prime)}) \mapsto \begin{cases}
        (i, r, r^\prime) & \text{if } j + 2^{k} < |T| \,,\\
        (i, r, 0)     & \mathrm{otherwise}.
      \end{cases}\right)$\label{alg:isa_window}
  }
\end{algorithm2e}
\SetAlgorithmName{Example}{example}{List of Examples}
\begin{algorithm2e}[t]
  \def\C#1{\texttt{#1}}
  \caption{Example of prefix doubling using the inverse suffix array in Thrill.\label{alg:generic_prefix_doubling_isa_example}}
  $T = \arr{\C{b},\C{d},\C{a},\C{c},\C{b},\C{d},\C{a},\C{c},\C{b} }$\;
  $S = \arr{ (0,\C{b},\C{d}), (1,\C{d},\C{a}), (2,\C{a},\C{c}), (3,\C{c},\C{b}), (4,\C{b},\C{d}), (5,\C{d},\C{a}),
             (6,\C{a},\C{c}), (7,\C{c},\C{b}), (8,\C{b},\C{\$}) }$
             \Rem{\ref{alg:prefix_doubling_generic}.\ref{alg:generic_initial_tuples}}
  $\mathbf{k = 1}$\Rem{\ref{alg:prefix_doubling_generic}.\ref{alg:generic_loop}}
  $S = \arr{ (2,\C{a},\C{c}), (6,\C{a},\C{c}), (8,\C{b},\C{\$}), (0,\C{b},\C{d}), (4,\C{b},\C{d}), (3,\C{c},\C{b}),
             (7,\C{c},\C{b}), (1,\C{d},\C{a}), (5,\C{d},\C{a}) }$
  \Rem{\ref{alg:prefix_doubling_generic}.\ref{alg:generic_sort_name_pair}}
  $N = \arr{ (2, 0), (6, 0), (8, 2), (0, 3), (4, 0), (3, 5), (7, 0), (1, 7), (5, 0) }$
  \Rem{\ref{alg:prefix_doubling_generic}.\ref{alg:generic_map_name}}
  \textbf{4 items with rank 0}\Rem{\ref{alg:prefix_doubling_generic}.\ref{alg:generic_finished}}
  $N = \arr{ (2, 0), (6, 0), (8, 2), (0, 3), (4, 3), (3, 5), (7, 5), (1, 7), (5, 7) }$
  \Rem{\ref{alg:prefix_doubling_generic}.\ref{alg:generic_new_name}}
  $N = \arr{ (0, 3), (1, 7), (2, 0), (3, 5), (4, 3), (5, 7), (6, 0), (7, 5), (8, 2) }$
  \Rem{\ref{alg:prefix_doubling_isa}.\ref{alg:isa_sort}}
  $S = \arr{ (0, 3, 0), (1, 7, 5), (2, 0, 3), (3, 5, 7), (4, 3, 0), (5, 7, 5), (6, 0, 2), (7, 5, 0), (8, 2, 0) }$
  \Rem{\ref{alg:prefix_doubling_isa}.\ref{alg:isa_window}}
  $\mathbf{k = 2}$\Rem{\ref{alg:prefix_doubling_generic}.\ref{alg:generic_loop}}
  $S = \arr{ (6, 0, 2), (2, 0, 3), (8, 2, 0), (0, 3, 0), (4, 3, 0), (7, 5, 0), (3, 5, 7), (1, 7, 5), (5, 7, 5) }$
  \Rem{\ref{alg:prefix_doubling_generic}.\ref{alg:generic_sort_name_pair}}
  $N = \arr{ (6, 0), (2, 1), (8, 2), (0, 3), (4, 0), (7, 5), (3, 6), (1, 7), (5, 0) }$
  \Rem{\ref{alg:prefix_doubling_generic}.\ref{alg:generic_map_name}}
  \textbf{2 items with rank 0}\Rem{\ref{alg:prefix_doubling_generic}.\ref{alg:generic_finished}}
  $N = \arr{ (6, 0), (2, 1), (8, 2), (0, 3), (4, 3), (7, 5), (3, 6), (1, 7), (5, 7) }$
  \Rem{\ref{alg:prefix_doubling_generic}.\ref{alg:generic_new_name}}
  $N = \arr{ (0, 3), (1, 7), (2, 1), (3, 6), (4, 3), (5, 7), (6, 0), (7, 5), (8, 2) }$
  \Rem{\ref{alg:prefix_doubling_isa}.\ref{alg:isa_sort}}
  $S = \arr{ (0, 3, 3), (1, 7, 7), (2, 1, 0), (3, 6, 5), (4, 3, 2), (5, 7, 0), (6, 0, 0), (7, 5, 0), (8, 2, 0) }$
  \Rem{\ref{alg:prefix_doubling_isa}.\ref{alg:isa_window}}
  $\mathbf{k = 3}$\Rem{\ref{alg:prefix_doubling_generic}.\ref{alg:generic_loop}}
  $S = \arr{ (6, 0, 0), (2, 1, 0), (8, 2, 0), (4, 3, 2), (0, 3, 3), (7, 5, 0), (3, 6, 5),  (5, 7, 0), (1, 7, 7) }$
  \Rem{\ref{alg:prefix_doubling_generic}.\ref{alg:generic_sort_name_pair}}
  $N = \arr{ (6, 0), (2, 1), (8, 2), (4, 3), (0, 4), (7, 5), (3, 6), (5, 7), (1, 8) }$
  \Rem{\ref{alg:prefix_doubling_generic}.\ref{alg:generic_map_name}}
  \textbf{1 item with rank 0}\Rem{\ref{alg:prefix_doubling_generic}.\ref{alg:generic_finished}}
  \textbf{Result:} $\arr{ 6, 2, 8, 4, 0, 7, 3, 5, 1 }$\Rem{\ref{alg:prefix_doubling_generic}.\ref{alg:generic_return}}
\end{algorithm2e}
\SetAlgorithmName{Algorithm}{algorithm}{List of Algorithms}

\subsubsection{Prefix Doubling with Discarding.}

In the algorithms described above, we always sort and consider all suffixes (name pairs) for the following renaming.
Even though some of them are already at their correct position, i.e., have a unique name.
Now, we present an algorithm which extends the prefix doubling algorithm using sorting such that only the following suffixes are sorted:
suffixes that do not yet have a unique name and suffixes that have an unique name but are required to compute a name pair (for a suffix that does not yet have a unique name).
All other suffixes are \emph{discarded} and are not considered for the computation anymore.
This technique has also been considered for external memory suffix array construction~\cite{dementiev08better}.

\begin{algorithm2e}[t]
  \caption{Prefix Doubling with Discarding.\label{alg:prefix_doubling_discarding}}
  \Function{PrefixDoublingDiscarding($T\in\DIA{\Sigma}$)}{
    $S:= T.\Window_2((i, \arr{t_0, t_1}) \mapsto (i, t_0, t_1) )$\Rem{Create initial triples $(i, T[i], T[i + 1])$.}\label{alg:discarding_init_name_pairs}
    $S:= S.\Sort((i,r_0, r_1) \text{ by } (r_0, r_1))$\Rem{Sort triples by name pairs.}
    $N:=S.\FlatWindow_2((i, \arr{a, b}) \mapsto \CmpName{i, a, b})$\Rem{Map names to 0 or $i$.}
    $N:=N.\PrefixSum(((i, r), (i^\prime, r^\prime))\mapsto (i^\prime, \max{(r, r^\prime)}))$\Rem{Calculate initial names.}
    \For{$k:= 1 \KwSty{ to } \lceil\log_2 |T|\rceil$}{
      $P:= N.\FlatWindow_3((i, \arr{a, b, c})\mapsto \Unique(a, b, c, i))$\Rem{Compute states of items.}
      $P:= \Union(P, U).\Sort(<^{k}_{\mathrm{op}})$ \Rem{Concatenate undiscarded items and sort them.}
      $P:= P.\FlatWindow_3((i, \arr{a, b, c})\mapsto \Discard(i, a, b, c, k))$\Rem{Compute new name}
      $D^\prime:= P.\Filter((i,r_0,r_1,s)) \mapsto (s = \texttt{d})$\Rem{pairs and update state. Then find and}
      $D:=\Union(D, D^\prime).\Map((i,r_0,r_1,s) \mapsto (i, a.r_0)))$\Rem{store newly discarded items.}
      $U^\prime := P.\Filter((i,r_0,r_1,s)\mapsto (s = \texttt{u}))$\Rem{Separate the already unique items and the}
      $U:=U^\prime.\Map((i,r_0,r_1,s) \mapsto (i, r_0, s))$\Rem{items that still need to be sorted. Former}
      $I^\prime := P.\Filter((i,r_0,r_1,s) \mapsto (s = \texttt{n}))$\Rem{are only needed to compute the name pairs}
      $I:=I^\prime.\Map((i,r_0,r_1,s) \mapsto (i, r_0, r_1))$\Rem{and stored in $U$. Latter are stored in $I$.}
      \If{$I.\Size()=0$}{\label{alg:discarding_all_names_unique}
        \textbf{return} $D.\Sort((i,r) \text{ by } r).\Map((i,r) \mapsto i)$\Rem{If all items are unique return \SA.}\label{alg:discarding_finished}
      }
      $M:= I.\FlatWindow_2((i, \arr{a, b}) \mapsto \text{NameDiscarding}(i, a, b))$\Rem{Form names that}\label{alg:discarding_begin_new_names}
      $M:=M.\PrefixSum(((i,r_0,r_1,r_2), (i^\prime,r_0^\prime,r_1^\prime,r_2^\prime))\mapsto (i^\prime,\max(r_0^\prime, r_0^\prime),\max(r_1^\prime, r_1^\prime), r_2^\prime))$\;
      $N:=M.\Map((i,r_0,r_1,r_2) \mapsto (i, r_2 + (r_1 - r_0)))$ \Rem{comply with the old names.}\label{alg:discarding_begin_end_names}
    }
  }
\end{algorithm2e}

\begin{algorithm2e}[p]
  \caption{Prefix Doubling with Discarding (Additional Functions)\label{alg:prefix_doubling_discarding_helper}}
  \Function{Unique($j\in\mathbb{N}_0$, $(i,r), (i^\prime,r^\prime),(i^{\prime\prime},r^{\prime\prime})\in N$)}{
    \If{$j = 0$}{
      \textbf{emit}~$\begin{cases} (i, r, \texttt{u}) & \text{if } r \neq r^\prime,\Rem{First item is unique} \\ (i, r, \texttt{n}) & \text{otherwise.}\Rem{if its ranks differ from its successor.} \end{cases}$ 
    }
    \ElseIf{$j + 2 = l$}{
      \textbf{emit}~$\begin{cases} (i^{\prime\prime}, r^{\prime\prime}, \texttt{u}) & \text{if } r^\prime \neq r^{\prime\prime},\Rem{Final item is unique} \\ (i^{\prime\prime}, r^{\prime\prime}, \texttt{n}) & \text{otherwise.}\Rem{if its ranks differs from its precursor.} \end{cases}$ 
    }
    \textbf{emit}~$\begin{cases} (i^\prime, r^\prime, \texttt{u}) & \text{if } r \neq r^\prime \text{ and } r^\prime \neq r^{\prime\prime},\Rem{An item is} \\ (i^\prime, r^\prime, \texttt{n}) & \text{otherwise.}\Rem{unique if its ranks are unique.} \end{cases}$
  }

  \Function{NPairs($j\in\mathbb{N}_0$, $(i,r,s), (i^\prime,r^\prime,s^\prime),(i^{\prime\prime},r^{\prime\prime},s^{\prime\prime})\in P$, $k\in\mathbb{N}_0$)}{\label{alg:npairs_begin}
    \If{$j = 0$}{
      \textbf{emit}~$\begin{cases} (i,r,0,\texttt{d}) & \text{if } s = \texttt{u},\Rem{The first two items can be discarded} \\ (i^\prime,r^\prime,0,\texttt{d}) & \text{if } s^\prime = \texttt{u}.\Rem{if they are unique. Emit $\leq 2$ items.} \end{cases}$
    }
    \ElseIf{$j + 2 = l$}{
      \If{$s^\prime = \mathtt{n}$}{
        \textbf{emit}~$\begin{cases} (i^\prime, r^\prime, r^{\prime\prime}, \texttt{n}) & \text{if } i^\prime+2^k=i^{\prime\prime},\Rem{If the last two items of the} \\ (i^\prime, r^\prime, 0, \texttt{n}) & \text{otherwise.}\Rem{\DIA{} are undecided, then we need} \end{cases}$ \;
      }
      \If{$s^{\prime\prime} = \mathtt{n}$}{
        \textbf{emit}~$(i^{\prime\prime},r^{\prime\prime},0,\mathtt{n})$\Rem{to fuse the ranks required for renaming.}
      }
    }
    \If{$s = \mathtt{n}$}{
      \textbf{emit}~$\begin{cases} (i, r, r^\prime, \texttt{n}) & \text{if } i+2^k=i^\prime,\Rem{The ranks for renaming are} \\ (i, r, 0, \texttt{n}) & \text{otherwise.}\Rem{consecutive and fused accordingly.} \end{cases}$ \;
    }
    \If{$s^{\prime\prime} = \mathtt{u}$}{
      \textbf{emit}~$\begin{cases} (i^{\prime\prime}, r^{\prime\prime}, 0, \texttt{d}) & \text{if } s =\mathtt{u} \text{ or } s^\prime =\mathtt{u},\Rem{Unique items are dis-} \\ (i^{\prime\prime}, r^{\prime\prime}, 0, \texttt{u}) & \text{otherwise.}\Rem{carded if uncalled-for in future renaming.} \end{cases}$
    }
  }\label{alg:npairs_end}

  \Function{NameDiscarding($j\in\mathbb{N}_0$, $l\in\mathbb{N}_0$, $(i,r_0,r_1),(i^\prime,r_0^\prime,r_1^\prime)\in I$)}{
    \If{$j = 0$}{
      \textbf{emit}~$(i, 1, 1, r_0)$ \Rem{The new names must comply with the old ones.}
    }
    \textbf{emit}~$\begin{cases} (i^\prime, j + 2, j + 2, r_0^\prime) & \text{if } r_0 \neq r_0^\prime \text{ and } r_1 \neq r_1^\prime,\Rem{The first rank de-} \\ (i^\prime, 1,j+2,r_0^\prime) & \text{else if } r_0 = r_0^\prime,\Rem{termines the group and new}\\ (i^\prime, 1,1,r_0^\prime) & \text{otherwise}\,.\Rem{names are consistent within groups.} \end{cases}$ 
  }
\end{algorithm2e}

Initially, Algorithm~\ref{alg:prefix_doubling_discarding} behaves like the generic prefix doubling algorithm (see Figure~\ref{fig:dataflow-pd-dis} for the data-flow graph).
We compute name pairs for consecutive text positions (line~\ref{alg:discarding_init_name_pairs}) and compute the names for all suffixes the same way we do in the generic algorithm.
Next, we add a \emph{state} to the triples $(i, r_1, r_2)$, i.e., creating $4$-tuples $(i, r_1, r_2, s)$, indicating whether a name pair is \emph{unique} (\texttt{u}) or \emph{not unique} (\texttt{n}) (see function \Unique, Algorithm~\ref{alg:prefix_doubling_discarding_helper}).
All $4$-tuples that are unique do not need a new name but they may still be required to compute the new name of another suffix.
Hence we add a third state, a $4$-tuple that is unique gets the state \emph{discarded} (\texttt{d}) if it is not required for the computation of a different name.
Those tuples can easily be identified by looking at three consecutive tuples after they have been sorted using the less-comparator described in Equation~\ref{eq:less_comparator}.
Let $a = (i, r_1, r_2, s), b = (i^\prime, r_1^\prime, r_2^\prime, s^\prime)$ and $c = (i^{\prime\prime}, r_1^{\prime\prime}, r_2^{\prime\prime}, s^{\prime\prime})$ be three continuous tuples with $s^{\prime\prime}$ being unique. If either $s$ or $s^\prime$ is unique, then $c$ can be discarded because both $a$ and $b$ will get a unique name pair during this iteration.
Otherwise (if $s$ and $s^\prime$ are not unique) then $c$ cannot be discarded as $a$ will not get a unique name pair during this iteration and we require the name of $c$ during the next iteration to compute the name pair (see Function \Discard, Algorithm~\ref{alg:prefix_doubling_discarding_helper}, Lines~\ref{alg:npairs_begin}--\ref{alg:npairs_end}).
While computing the final state we also create the new name pairs required for the new name if the state is not unique, as otherwise the name is final.

\begin{figure}
  \centering
  \begin{tikzpicture}[scale=0.4, yscale=0.8]
    \node (1) at (400bp,1746bp) [] {$T$};
    \node (2) at (400bp,1674bp) [DOp] {$S := T.\Window_2$};
    \node (3) at (400bp,1602bp) [DOp] {$S := S.\Sort$};
    \node (4) at (400bp,1530bp) [DOp] {$N := S.\FlatWindow_2$};
    \node (5) at (400bp,1458bp) [DOp] {$N := N.\PrefixSum$};
    \node (6) at (400bp,1386bp) [DOp] {$P := N.\FlatWindow_3$};
    \node (7) at (400bp,1314bp) [DOp] {$P := P.\Sort$};
    \node (8) at (200bp,1242bp) [Action] {$P.\Size$};
    \node (9) at (400bp,1242bp) [DOp] {$P := P.\FlatWindow_3$};
    \node (10) at (200bp,1170bp) [LOp] {$D' := P.\Filter$};
    \node (11) at (200bp,1098bp) [LOp] {$D' := D'.\Map$};
    \node (12) at (400bp,1170bp) [LOp] {$U' = P.\Filter$};
    \node (13) at (400bp,1098bp) [LOp] {$U := U'.\Map$};
    \node (14) at (600bp,1170bp) [LOp] {$I' := P.\Filter$};
    \node (15) at (600bp,1098bp) [LOp] {$I := I'.\Map$};
    \node (16) at (600bp,1026bp) [DOp] {$I := I.\Sort$};
    \node (17) at (200bp,1026bp) [DOp] {$\Cache$};
    \node (18) at (600bp,954bp) [Action] {$I.\Size$};
    \node (19) at (400bp,954bp) [DOp] {$M := I.\FlatWindow_2$};
    \node (20) at (400bp,882bp) [DOp] {$M := \PrefixSum$};
    \node (21) at (400bp,810bp) [LOp] {$N := M.\Map$};
    \node (22) at (400bp,738bp) [DOp] {$P := N.\FlatWindow_3$};
    \node (23) at (400bp,666bp) [DOp] {$P := \Union(U, P)$};
    \node (24) at (400bp,594bp) [DOp] {$P := P.\Sort$};
    \node (25) at (200bp,522bp) [Action] {$P.\Size$};
    \node (26) at (400bp,522bp) [DOp] {$P := P.\FlatWindow_3$};
    \node (27) at (200bp,450bp) [DOp] {$D' := P.\Filter$};
    \node (28) at (200bp,378bp) [LOp] {$D' := D'.\Map$};
    \node (29) at (400bp,450bp) [LOp] {$U' := P.\Filter$};
    \node (30) at (400bp,378bp) [LOp] {$U := U'.\Map$};
    \node (31) at (600bp,450bp) [LOp] {$I' := P.\Filter$};
    \node (32) at (600bp,378bp) [LOp] {$I := I'.\Map$};
    \node (33) at (600bp,306bp) [DOp] {$I := I.\Sort$};
    \node (34) at (200bp,306bp) [DOp] {$\Cache$};
    \node (35) at (600bp,234bp) [Action] {$I.\Size$};
    \node (36) at (200bp,234bp) [DOp] {$D := \Union(D, D')$};
    \node (37) at (300bp,162bp) [DOp] {$D.\Sort.\Map$};
    \node (39) at (450bp,162bp) [] {$\SA_T$};
    \draw [->] (1) -- (2);
    \draw [->] (2) -- (3);
    \draw [->] (3) -- (4);
    \draw [->] (4) -- (5);
    \draw [->] (5) -- (6);
    \draw [->] (6) -- (7);
    \draw [->] (7) -- (8);
    \draw [->] (7) -- (9);
    \draw [->] (9) -- (10);
    \draw [->] (9) -- (12);
    \draw [->] (9) -- (14);
    \draw [->] (10) -- (11);
    \draw [->] (11) -- (17);
    \draw [->] (12) -- (13);
    \draw [->] (13) to[out=270,in=0] +(-100bp,-80bp) to[out=180,in=180,looseness=0.7] (23);
    \draw [->] (14) -- (15);
    \draw [->] (15) -- (16);
    \draw [->] (16) -- (18);
    \draw [->] (16) -- (19);
    \draw [->,overlay] (17) to[out=270,in=160,looseness=0.6] (36);
    \draw [->] (19) -- (20);
    \draw [->] (20) -- (21);
    \draw [->] (21) -- (22);
    \draw [->] (22) -- (23);
    \draw [->] (23) -- (24);
    \draw [->] (24) -- (25);
    \draw [->] (24) -- (26);
    \draw [->] (26) -- (27);
    \draw [->] (26) -- (29);
    \draw [->] (26) -- (31);
    \draw [->] (27) -- (28);
    \draw [->] (28) -- (34);
    \draw [->] (29) -- (30);
    \draw [->] (31) -- (32);
    \draw [->] (32) -- (33);
    \draw [->] (33) -- (35);
    \draw [->] (34) -- (36);
    \draw [->] (36) -- (37);
    \draw [->] (37) -- (39);
  \end{tikzpicture}
  \caption{DIA data-flow graph of two iterations of prefix doubling with discarding.}\label{fig:dataflow-pd-dis}
\end{figure}

Since we do not consider all tuples during the course of Algorithm~\ref{alg:prefix_doubling_discarding} we need to change the renaming based on the name pairs.
Up to now, we were able to give names starting at $0$ and continue based on the (preliminary) position in SA.
If we discard tuples this approach is not feasible any more as we need to consider the names of already discarded tuples.
During the $k$-th iteration, all suffixes that do not have a unique name form consecutive intervals in \SA.
Within these intervals all suffixes that cannot be distinguished by their first $2^k$ characters share the same name.
These names are extended, i.e., increased such that the new name is always at least as great as the previous name and greater than the rank of the first preceding suffix that can be distinguished using the first $2^k$ characters of the suffixes (lines~\ref{alg:discarding_begin_new_names}--\ref{alg:discarding_begin_end_names}).
At the beginning of the next iteration we add all unique names to the new names and check if they can be discarded.
As soon as all names are unique (line~\ref{alg:discarding_all_names_unique}) we know that all names have been discarded and can compute \SA{} by sorting the discarded tuples by their names (line~\ref{alg:discarding_finished}).

\subsubsection{Prefix Quadrupling.}
In the prefix doubling algorithms described above, during the $k$-th doubling step, we consider substrings of length $2^{k}$.
This can be generalized to substrings of length $a^{k}$ for any $a\in\mathbb{N}_0$ with $a>1$.
The prefix doubling algorithms using sorting are I/O optimal for $5$-tuples in external memory and in practice using $4$-tuples, i.e., prefix quadrupling has the advantage that less memory is required for storing the tuples and that the I/O-volume is just $1.5\%$ worse compared to prefix quintupling~\cite{dementiev08better}.
The change within the algorithms can be kept to a minimum as we just require rank quadruples instead of rank pairs.
Also, the comparison and the computation of the new names have to be adapted accordingly.

\subsection{Difference Cover Algorithms -- DC3 and DC7, aka skew3 and skew7}
\label{ssec:difference_cover_algorithms}
In 2003, the DC3 aka skew3 suffix sorting algorithm and its generalization, DC$X$ and skew$X$, was proposed by Kärkkäinen, Sanders, and Burkhardt~\cite{karkkainen2003simple,karkkainen2006linear}.
They employ recursion on a subset of the suffixes to reach linear running time in the sequential RAM model, which translates to sorting complexity in the external memory and PRAM models.
While the reference implementation by the authors is in the sequential RAM model, the algorithms were later implemented for external memory~\cite{dementiev08better,weese06entwurf}, and DC3 was implemented for distributed memory using MPI~\cite{kulla2007scalable}.

The DC$X$ algorithms are based on scanning, sorting, and merging, and hence are asymptotically optimal in many models provided optimal theoretical base algorithms.
As Thrill supplies all of these base algorithms as scalable distributed algorithmic primitives, implementing DC$X$ is a natural choice.

The key notion of DC$X$ is to recursively calculate the ranks of suffixes in only a \emph{difference cover} of the original text. A set $D \subseteq \mathbb{N}_0$ is a difference cover for $n \in \mathbb{N}_0$, if $\{ (i - j) \bmod n \mid i,j  \in D \} = \{ 0, \ldots, n-1 \}$. Examples of difference covers are $D_3 = \{ 1, 2 \}$ for $n = 3$, $D_7 = \{ 0, 1, 3 \}$ for $n = 7$, and $D_{13} =  \{ 0, 1, 3, 9 \}$ for $n = 13$. In general, a difference cover of size $\mathcal{O}(\sqrt{n})$ can be calculated for any $n$ in $\mathcal{O}(\sqrt{n})$ time~\cite{karkkainen2006linear}.

The broad steps of the DC3 algorithm are the following:
\begin{enumerate}
\item Calculate ranks for all suffixes starting at positions $i$ in the difference cover $D_3 = \{ 1, 2 \}$.
  This is done by sorting the triples $(T[i],T[i+1],T[i+2])$ for $i \in D_3$, calculating lexicographic names, and recursively calling a suffix sorting algorithm on a reduced string of size $\frac{2}{3} n$ if necessary.
  The result of step 1 are two arrays, $R_1$ and $R_2$, containing the ranks of suffixes $i = 1 \bmod 3$ and $i = 2 \bmod 3$.

\item Scan text $T$, $R_1$, and $R_2$ to generate three arrays: $S_0$, $S_1$, and $S_2$, where array $S_j$ contains one tuple for each suffix $i$ with $i = j \bmod 3$.
  The arrays store tuples containing the two next ranks from $R_1$ and $R_2$ and all characters from $T$ up to the next ranks.
  This is exactly the information required such that the following merge step is able to deduce the suffix array.

\item Sort $S_0$, $S_1$, and $S_2$ and merge them using a custom comparison function which compares the suffixes represented in the tuples using characters and ranks.
  Only a constant number of characters and ranks need to be accessed in each comparison.
  Output the suffix array using the indices stored in tuples.

\end{enumerate}

\begin{algorithm2e}
  \SetKwFunction{MakeTriples}{MakeTriples}
  \def\CmpLine[#1&#2&#3&#4&#5&#6]{
    \parbox[b]{11ex}{#1}\parbox[b]{14ex}{#2}%
    if \parbox[b]{21.5ex}{#3}\parbox[b]{6ex}{#4} \parbox[b]{22.5ex}{#5}\parbox[b]{6ex}{#6,}%
  }
  \caption{DC3 Algorithm in Thrill.\label{alg:dc3}}
  \Function{DC3($T \in \DIA{\Sigma}$)}{
    $T_3 := T.\FlatWindow_3((i, \arr{c_0,c_1,c_2}) \mapsto \MakeTriples(i,c_0,c_1,c_2))$ \;
    \KwSty{with} \Function{\MakeTriples($i \in \mathbb{N}_0$, $c_0,c_1,c_2 \in \Sigma$)}{
      \lIf(\Remi{Make triples $i \in D_3$.}){$i \neq 0 \bmod 3$}{
        \KwSty{emit} $(i,c_0,c_1,c_2)$
      }
    }
    $S := T_3.\Sort((i,c_0,c_1,c_2) \text{ by } (c_0,c_1,c_2))$ \Rem{Sort triples lexicographically.}
    $I_S := S.\Map((i,c_0,c_1,c_2) \mapsto i)$ \Rem{Extract sorted indices.}
    $N' := S.\FlatWindow_2((i, \arr{p_0,p_1}) \mapsto \operatorname{CmpTriple}(i,p_0,p_1))$ \Rem{Compare triples.}
    \KwSty{with} \Function{CmpTriple($i \in \mathbb{N}_0$, $p_0 = (c_0,c_1,c_2)$, $p_1 = (c'_0,c'_1,c'_2)$) \Remi{Emit one}}{
      \lIf(\Remi{sentinel for index 0, and 0 or 1}){$i = 0$}{
        \KwSty{emit} $0$
      }
      \KwSty{emit} $\KwIfThenElse{(c_0,c_1,c_2) = (c'_0,c'_1,c'_2)}{0}{1}$ \Rem{depending on previous tuple.}
    }
    $N := N'.\PrefixSum()$ \Rem{Use prefix sum to calculate names.}
    $n_{\text{sub}} = \lceil{2 |T| / 3}\rceil$, \quad $n_{\text{mod1}} = \lceil{|T| / 3}\rceil$ \Rem{Size of recursive problem and mod 1 part of $T_R$}
    \uIf(\Remi{If duplicate names exist, sort names back to}){$N.\Max() + 1 = n_{\text{sub}}$}{
      $T'_R := \Zip(\arr{ I_S, N }, (i, n) \mapsto (i,n)).\Sort((i,n) \text{ by } (i \bmod 3, i \text{ div } 3))$ \Rem{string order}
      $\SA_R := \operatorname{DC3}(T'_R.\Map( (i,n) \mapsto n ))$ \Rem{as $T_1 \oplus T_2$ and call suffix sorter.}
      $I'_R := \SA_R.\ZipWithIndex((i,r) \mapsto (i,r))$ \Rem{Invert resulting suffix array, but}
      $I_R := I'_R.\Sort((i,r) \text{ by } (i \bmod n_{\text{mod1}}, i))$ \Rem{interleave $\ISA$ for better locality}
      $R_1 := I_R.\Filter((i,r) \mapsto i < n_{\text{mod1}}).\Map((i,r) \mapsto r + 1)$ \Rem{after separating $\ISA$}
      $R_2 := I_R.\Filter((i,r) \mapsto i \geq n_{\text{mod1}}).\Map((i,r) \mapsto r + 1)$ \Rem{ into $R_1$ and $R_2$.}
    }
    \Else(\Remi{Else, if all names/triples are unique, then $I_S$ is already the suffix array.}){
      $R := I_S.\ZipWithIndex((i,r) \mapsto (i,r))$ \Rem{Invert it to get $\ISA$, but}
      $I_R := R.\Sort( (i,r) \text{ by } (i \text{ div } 3, i) )$ \Rem{interleave $\ISA$ for better locality}
      $R_1 := I_R.\Filter((i,r) \mapsto i = 1 \bmod 3).\Map((i,r) \mapsto r + 1)$ \Rem{after separating it}
      $R_2 := I_R.\Filter((i,r) \mapsto i = 2 \bmod 3).\Map((i,r) \mapsto r + 1)$ \Rem{into $R_1$ and $R_2$.}
    }
    $\hat{T}_3 := T.\FlatWindow_3((i, \arr{c_0,c_1,c_2}) \mapsto \MakeTriples(i,c_0,c_1,c_2))$ \;
    \KwSty{with} \Function{\MakeTriples($i \in \mathbb{N}_0$, $c_0,c_1,c_2 \in \Sigma$) \Remi{Prepare Zip with all}}{
      \lIf(\Remi{triples $i \notin D_3$.}){$i = 0 \bmod 3$}{
        \KwSty{emit} $(c_0,c_1,c_2)$
      }
    }
    $Z' := \Zip(\arr{ \hat{T}_3,R_1,R_2 }, ((c_0,c_1,c_2), r_1, r_2) \mapsto (c_0,c_1,c_2,r_1,r_2))$ \Rem{Pull chars and ranks}
    $Z := Z'.\Window_2( (i, \arr{ (z_1,z_2) }) \mapsto (i, z_1, z_2) )$ \Rem{using Zip from three arrays}
    $S'_0 := Z.\Map( (i, (c_0,c_1,c_2,r_1,r_2), (\bar{c}_0,\bar{c}_1,\bar{c}_2,\bar{r}_1,\bar{r}_2)) \mapsto (3 i + 0, c_0, c_1, r_1, r_2) )$ \Rem{to make}
    $S'_1 := Z.\Map( (i, (c_0,c_1,c_2,r_1,r_2), (\bar{c}_0,\bar{c}_1,\bar{c}_2,\bar{r}_1,\bar{r}_2)) \mapsto (3 i + 1, c_1, r_1, r_2) )$ \Rem{arrays of}
    $S'_2 := Z.\Map( (i, (c_0,c_1,c_2,r_1,r_2), (\bar{c}_0,\bar{c}_1,\bar{c}_2,\bar{r}_1,\bar{r}_2)) \mapsto (3 i + 2, c_2, r_2, \bar{c}_0, \bar{r}_1) )$\;
    $S_0 := S'_0.\Sort( (i, c_0, c_1, r_1, r_2) \text{ by } (c_0, r_1) )$ \Rem{representatives for each}
    $S_1 := S'_1.\Sort( (i, c_1, r_1, r_2) \text{ by } (r_1) )$ \Rem{suffix class.}
    $S_2 := S'_2.\Sort( (i, c_2, r_2, \bar{c}_0, \bar{r}_1) \text{ by } (r_2) )$ \;
    \Return $\Merge(\arr{ S_0,S_1,S_2 }, \operatorname{CompareDC3}).\Map((i,\ldots) \mapsto i)$ \Rem{Merge sorted}
    \KwSty{with} \Function{CompareDC3($z_1,z_2$) \Remi{representatives to deliver final suffix array.}}{
      \CmpLine[$(c_0,r_1)$     & $< (c'_0,r'_1)$                  & $z_1 = (i, c_0, c_1, r_1, r_2)$             & $\in S_0$, & $z_2 = (i', c'_0, c'_1, r'_1, r'_2)$             & $\in S_1$]\;
      \CmpLine[$(c_0,r_1)$     & $< (c'_1,r'_2)$                  & $z_1 = (i, c_0, c_1, r_1, r_2)$             & $\in S_0$, & $z_2 = (i', c'_1, r'_1, r'_2)$                   & $\in S_1$]\;
      \CmpLine[$(c_0,c_1,r_2)$ & $< (c'_2,\bar{c}'_0,\bar{r}'_1)$ & $z_1 = (i, c_0, c_1, r_1, r_2)$             & $\in S_0$, & $z_2 = (i', c'_2, r'_2, \bar{c}'_0, \bar{r}'_1)$ & $\in S_2$]\;
      \CmpLine[$(r_1)$         & $< (r'_1)$                       & $z_1 = (i, c_1, r_1, r_2)$                  & $\in S_1$, & $z_2 = (i', c'_1, r'_1, r'_2)$                   & $\in S_1$]\;
      \CmpLine[$(r_1)$         & $< (r'_2)$                       & $z_1 = (i, c_1, r_1, r_2)$                  & $\in S_1$, & $z_2 = (i', c'_2, r'_2, \bar{c}'_0, \bar{r}'_1)$ & $\in S_2$]\;
      \CmpLine[$(r_2)$         & $< (r'_2)$                       & $z_1 = (i, c_2, r_2, \bar{c}_0, \bar{r}_1)$ & $\in S_2$, & $z_2 = (i', c'_2, r'_2, \bar{c}'_0, \bar{r}'_1)$ & $\in S_2$]\;
      and symmetrically if $z_1 \in S_i, z_2 \in S_j \text{ with } i > j$ \,.
      }
  }
\end{algorithm2e}

\SetAlgorithmName{Example}{example}{List of Examples}
\begin{algorithm2e}[t]
  \def\C#1{\texttt{#1}}
  \caption{Example of DC3 Algorithm in Thrill.\label{alg:dc3 example}}
  $T = \arr{\C{d},\C{b},\C{a},\C{c},\C{b},\C{a},\C{c},\C{b},\C{d}}$\Rem{Example text $T$.}
  $T_3 = \arr{ (1,\C{b},\C{a},\C{c}), (2,\C{a},\C{c},\C{b}), (4,\C{b},\C{a},\C{c}), (5,\C{a},\C{c},\C{b}), (7,\C{b},\C{d},\C{\$}), (8,\C{d},\C{\$},\C{\$}) }$\Rem{Triples $i \in D_3$.}
  $S = \arr{ (2,\C{a},\C{c},\C{b}), (5,\C{a},\C{c},\C{b}), (1,\C{b},\C{a},\C{c}), (4,\C{b},\C{a},\C{c}), (7,\C{b},\C{d},\C{\$}), (8,\C{d},\C{\$},\C{\$}) }$\Rem{Sorted triples.}
  $I_S = \arr{ 2, 5, 1, 4, 7, 8 }$\Rem{Indexes extracted from sorted triples.}
  $N' = \arr{ 0, 0, 1, 0, 1 }$\Rem{0/1 indicators depending if triples are unequal or equal.}
  $N = \arr{ 0, 0, 1, 1, 2, 3 }$\Rem{Prefix sum of 0/1 indicators delivers lexicographic names.}
  $n_{\text{sub}} = 6$, $n_{\text{mod1}} = 3$\Rem{Calculate result size directly.}
  Condition $(N.\Max() + 1 = 4) \neq (6 = n_{\text{sub}})$, so follow recursion branch.\;
  $T''_R = \arr{ (2,0), (5,0), (1,1), (4,1), (7,2), (8,3) }$\Rem{Zip lexicographic names and their string}
  $T'_R = \arr{ (1,1), (4,1), (7,2), (2,0), (5,0), (8,3) }$\Rem{index, and sort them to string order}
  $T_R = \arr{ 1, 1, 2, 0, 0, 3 }$\Rem{to construct the recursive subproblem.}
  $\SA_R = \arr{ 3, 4, 0, 1, 2, 5 }$\Rem{Recursively calculate suffix array of $T_R$.}
  $I'_R = \arr{ (0,3), (1,4), (2,0), (3,1), (4,2), (5,5) }$\Rem{Add index positions to suffix array}
  $I_R = \arr{ (0,2), (3,0), (1,3), (4,1), (2,4), (5,5) }$\Rem{and sort into interleaved $R_1$ and $R_2$ ranks.}
  $R_1 = \arr{ 3, 4, 5 }$,\quad $R_2 = \arr{ 1, 2, 6 }$\Rem{Filter $R_1$ and $R_2$ from $I_R$.}
  $\hat{T}_3 = \arr{ (\C{d},\C{b},\C{a}), (\C{c},\C{b},\C{a}), (\C{c},\C{b},\C{d}) }$\Rem{Prepare triples $i \notin D_3$.}
  $Z' = \arr{ (\C{d},\C{b},\C{a},3,1), (\C{c},\C{b},\C{a},4,2), (\C{c},\C{b},\C{d},5,6) }$\Rem{Zip $\hat{T}_3$, $R_1$, $R_2$ to make arrays $S_i$.}
  $S'_0 = \arr{ (0,\C{d},\C{b},3,1), (3,\C{c},\C{b},4,2), (6,\C{c},\C{b},5,6) }$\Rem{Construct $(i, c_0, c_1, r_1, r_2) \in S_0$,}
  $S'_1 = \arr{ (1,\C{b},3,1), (4,\C{b},4,2), (7,\C{b},5,6) }$\Rem{$(i, c_1, r_1, r_2) \in S_1$, and}
  $S'_2 = \arr{ (2,\C{a},1,\C{c},4), (5,\C{a},2,\C{c},5), (8,\C{d},6,\C{\$},0) }$\Rem{$(i, c_2, r_2, \bar{c}_0, \bar{r}_1) \in S_2$}
  $S_0 = \arr{ (3,\C{c},\C{b},4,2), (6,\C{c},\C{b},5,6), (0,\C{d},\C{b},3,1) }$\Rem{as representatives of suffixes,}
  $S_1 = \arr{ (1,3,\C{b},1), (4,4,\C{b},2), (7,5,\C{b},6) }$\Rem{sort them among themselves}
  $S_2 = \arr{ (2,1,\C{a},\C{c},4), (5,2,\C{a},\C{c},5), (8,6,\C{d},\C{\$},0) }$\Rem{such that merging delivers}
  Result: $\arr{ 2, 5, 1, 4, 7, 3, 6, 8, 0 }$\Rem{the final suffix array.}
\end{algorithm2e}
\SetAlgorithmName{Algorithm}{algorithm}{List of Algorithms}

\begin{figure}
  \centering
  \begin{tikzpicture}[scale=0.4]
    \node (1) at (350bp,1386bp) {$T$};
    \node (2) at (350bp,1314bp) [DOp] {$T_7 := T.\FlatWindow_7$};
    \node (3) at (350bp,1242bp) [DOp] {$S := T_7.\Sort$};
    \node (4) at (350bp,1170bp) [LOp] {$I_s := S.\Map$};
    \node (5) at (350bp,1098bp) [DOp] {$\Cache$};
    \node (6) at (150bp,1170bp) [DOp] {$N' := S.\FlatWindow_2$};
    \node (7) at (150bp,1098bp) [DOp] {$N := N'.\PrefixSum$};
    \node (8) at (150bp,1026bp) [Action] {$N.\Max$};
    \node (10) at (350bp,1026bp) [DOp] {$R := I_S.\ZipWithIndex$};
    \node (11) at (350bp,954bp) [DOp] {$I_R := R.\Sort$};
    \node (12) at (150bp,882bp) [LOp] {$R'_0 := I_R.\Filter$};
    \node (13) at (150bp,810bp) [LOp] {$R_0 := R'_0.\Map$};
    \node (14) at (150bp,738bp) [LOp] {$\Collapse$};
    \node (15) at (350bp,882bp) [LOp] {$R'_1 := I_R.\Filter$};
    \node (16) at (350bp,810bp) [LOp] {$R_1 := R'_1.\Map$};
    \node (17) at (350bp,738bp) [LOp] {$\Collapse$};
    \node (18) at (550bp,882bp) [LOp] {$R'_3 := I_R.\Filter$};
    \node (19) at (550bp,810bp) [LOp] {$R_3 := R'_3.\Map$};
    \node (20) at (550bp,738bp) [LOp] {$\Collapse$};
    \node (21) at (650bp,954bp) [DOp] {$\hat{T}_7 := T.\FlatWindow_7$};
    \node (22) at (350bp,666bp) [DOp] {$Z' := \Zip(\arr{\hat{T}_7, R_0, R_1, R_3})$};
    \node (23) at (350bp,594bp) [DOp] {$Z := Z'.\Window_2$};
    \node (24) at (50bp,512bp) [LOp] {$S'_0 := Z.\Map$};
    \node (26) at (150bp,460bp) [LOp] {$S'_1 := Z.\Map$};
    \node (28) at (250bp,512bp) [LOp] {$S'_2 := Z.\Map$};
    \node (30) at (350bp,460bp) [LOp] {$S'_3 := Z.\Map$};
    \node (32) at (450bp,512bp) [LOp] {$S'_4 := Z.\Map$};
    \node (34) at (550bp,460bp) [LOp] {$S'_5 := Z.\Map$};
    \node (36) at (650bp,512bp) [LOp] {$S'_6 := Z.\Map$};
    \node (38) at (50bp,378bp) [DOp] {$S_0 := S'_0.\Sort$};
    \node (39) at (150bp,326bp) [DOp] {$S_1 := S'_1.\Sort$};
    \node (40) at (250bp,378bp) [DOp] {$S_2 := S'_2.\Sort$};
    \node (41) at (350bp,326bp) [DOp] {$S_3 := S'_3.\Sort$};
    \node (42) at (450bp,378bp) [DOp] {$S_4 := S'_4.\Sort$};
    \node (43) at (550bp,326bp) [DOp] {$S_5 := S'_5.\Sort$};
    \node (44) at (650bp,378bp) [DOp] {$S_6 := S'_6.\Sort$};
    \node (52) at (350bp,234bp) [DOp] {$\Merge(\arr{S_0,S_1,\ldots,S_6})$};
    \node (53) at (350bp,162bp) [] {$\SA_T$};
    \draw [->] (1) -- (2);
    \draw [->] (1) to[out=-20,in=90,looseness=1] ($(21) + (0,200bp)$) -- (21);
    \draw [->] (2) -- (3);
    \draw [->] (3) -- (4);
    \draw [->] (3) -- (6);
    \draw [->] (4) -- (5);
    \draw [->] (5) -- (10);
    \draw [->] (6) -- (7);
    \draw [->] (7) -- (8);
    \draw [->] (10) -- (11);
    \draw [->] (11) -- (12);
    \draw [->] (11) -- (15);
    \draw [->] (11) -- (18);
    \draw [->] (12) -- (13);
    \draw [->] (13) -- (14);
    \draw [->] (14) -- (22);
    \draw [->] (15) -- (16);
    \draw [->] (16) -- (17);
    \draw [->] (17) -- (22);
    \draw [->] (18) -- (19);
    \draw [->] (19) -- (20);
    \draw [->] (20) -- (22);
    \draw [->] (21) -- +(0,-150bp) to[out=270,in=0,looseness=1] (22);
    \draw [->] (22) -- (23);
    \draw [->] (23) to[out=190,in=90,looseness=0.5] (24);
    \draw [->] (23) to[out=200,in=90,looseness=1.6] (26);
    \draw [->] (23) to[out=230,in=90,looseness=1] (28);
    \draw [->] (23) to[out=270,in=90,looseness=1] (30);
    \draw [->] (23) to[out=310,in=90,looseness=1] (32);
    \draw [->] (23) to[out=340,in=90,looseness=1.6] (34);
    \draw [->] (23) to[out=350,in=90,looseness=0.5] (36);
    \draw [->] (24) -- (38);
    \draw [->] (26) -- (39);
    \draw [->] (28) -- (40);
    \draw [->] (30) -- (41);
    \draw [->] (32) -- (42);
    \draw [->] (34) -- (43);
    \draw [->] (36) -- (44);
    \draw [->] (38) -- +(0,-60bp) to[out=270,in=150,looseness=0.7] (52);
    \draw [->] (39) to[out=270,in=130,looseness=0.5] (52);
    \draw [->] (40) -- +(0,-40bp) to[out=270,in=110,looseness=1.4] (52);
    \draw [->] (41) to[out=270,in=90,looseness=1] (52);
    \draw [->] (42) -- +(0,-40bp) to[out=270,in=70,looseness=1.4] (52);
    \draw [->] (43) to[out=270,in=50,looseness=0.5] (52);
    \draw [->] (44) -- +(0,-60bp) to[out=270,in=30,looseness=0.7] (52);
    \draw [->] (52) -- (53);
  \end{tikzpicture}
  \caption{DIA data-flow graph of DC7 with no recursion.}\label{fig:dataflow-dc7}
\end{figure}
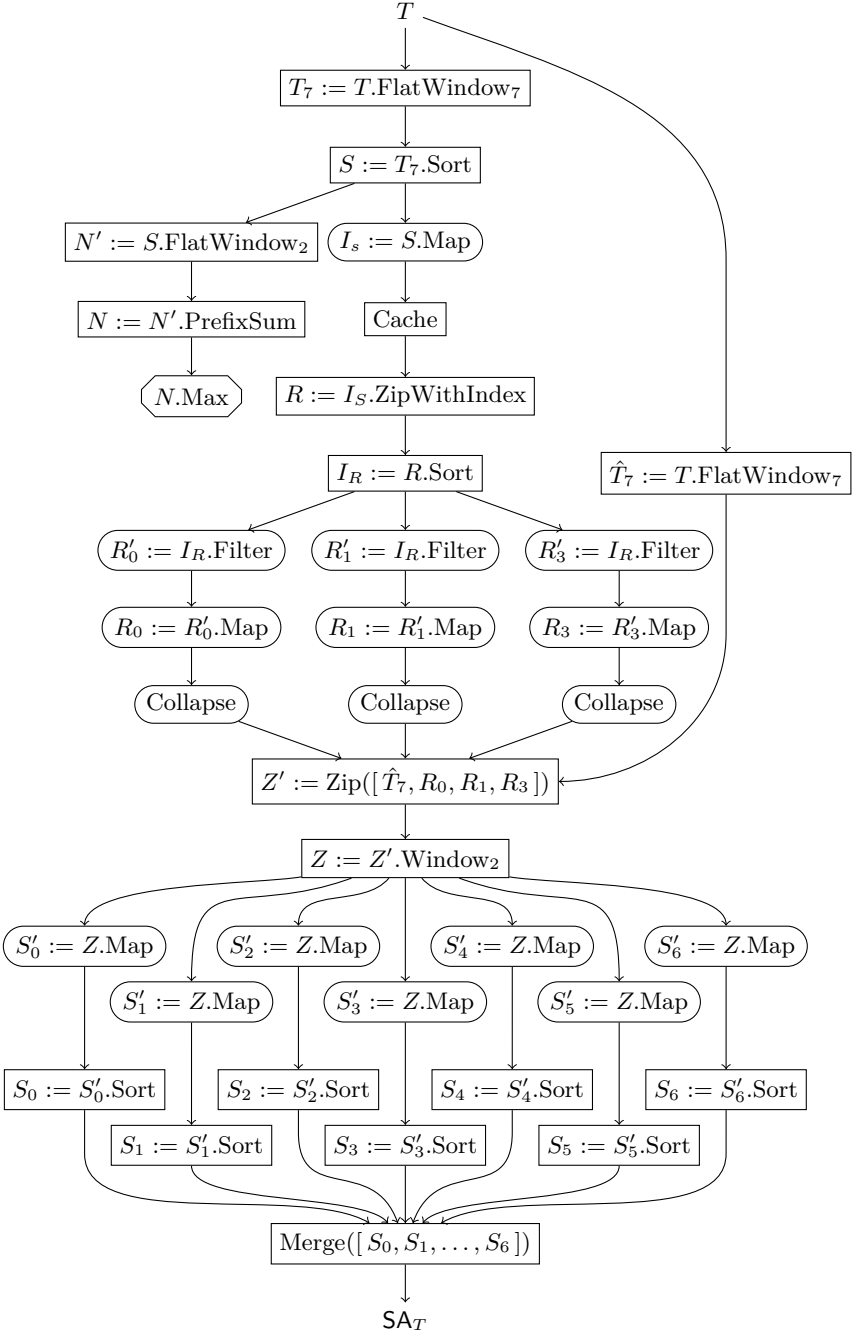

The first two steps of the difference cover suffix sorting algorithms can be seen as preparation for the final merge in Step 3.
Step 1 delivers ranks for all suffixes $i \in D_3$ in $R_1$ and $R_2$.
In Step 2 tuples are created in $S_0$, $S_1$, and $S_2$ which are constructed from the recursively calculated ranks and characters from the text.
The tuples are designed such that the comparison function can fully determine the final suffix array.
The complete DC3 implementation in Thrill algorithm code is shown as Algorithm~\ref{alg:dc3}, and Example~\ref{alg:dc3 example} shows the transcript of a run with the text $T=\texttt{dbacbacbd}$.
Figure~\ref{fig:dataflow-dc7} shows the dataflow graph of DC7 instead of DC3, which is slightly more complex but shows the algorithmic structure better.
In the algorithm code we omitted some details on padding and sentinels needed for inputs that are not a multiple of the difference cover size.

Goal of Lines 2--24 is to calculate $R_1$ and $R_2$ (step 1). This is done by performing the following steps:
\begin{enumerate}
\item Scan the text $T$ using a $\FlatWindow$ operation and create triples $(i,c_0,c_1,c_2)$ for all indices $i$ in the difference cover $D_3 = \{1,2\}$ (lines~2--4).
\item Sort the triples as $S$, scan $S$ and use a prefix sum to calculate lexicographic names $N$ (lines~5--11).
  The lexicographic names are constructed in the prefix sum from $0$ and $1$ indicators.
  The value $0$ is used if two lexicographic consecutive triples are equal, which means they are assigned the same lexicographic name; the value $1$ increments the name in the prefix sum and assigns the unequal triple a new name.
\item Check if all lexicographic names are different by comparing the highest lexicographic name against the maximum possible (lines~12--13)
\item If all lexicographic names are different, then $I_S$, which contains the indexes of $S$, is already the suffix array of the suffixes in $D_3$ (lines~21--24).
  Hence, $R_1$ and $R_2$ can be created directly: the suffix array $I_S$ only needs to be inverted and split by $\bmod 3$.
  Because Thrill's $\Filter$ operation is performed locally, we interleave the future $R_1$ and $R_2$ parts using the $\Sort$ operation such that the two arrays are balanced on the distributed system after the $\Filter$.
\item Otherwise, prepare a recursive subproblem $T_R$ to calculate the ranks.
  \begin{enumerate}
  \item Sort the lexicographic names back into string order such that $T_R = T_1 \oplus T_2$ where $\oplus$ is string concatenation (line~14).
    $T_1$ represents the complete text $T$ using the lexicographic names of all triples $i = 1 \bmod 3$, and $T_2$ is another complete copy of $T$ with triples $i = 2 \bmod 3$. By replacing the triples with lexicographic names, the original text is reduced by $\frac{2}{3}$.
  \item Recursively call any suffix sorting algorithm (e.g. DC3) on $T_R$ (line~15).
  \item Invert the permutation $\SA_R$ to gain ranks $R_1$ and $R_2$ of triples of $T$ in $D_3$, again interleave $\ISA_R$ such that $R_1$ and $R_2$ are balanced on the workers after the $\Filter$.
  \end{enumerate}
\end{enumerate}

With $R_1$ and $R_2$ from Step 1 (lines~2--24), the objective of Step 2 is to create $S_0$, $S_1$, and $S_2$ in Lines~25--35.
Each suffix $i$ has exactly one representative in the array $S_j$ where $j = i \bmod 3$.
Its representative contains the recursively calculated ranks of the two following suffixes in the difference cover from $R_1$ and $R_2$, and the characters $T[i],T[i+1],T[i+2],\ldots$ up to (but excluding) the next known rank.

For DC3 these are $R_1[\frac{i}{3}]$, $R_2[\frac{i}{3}]$, $T[i]$, and $T[i+1]$ for a suffix $i = 0 \bmod 3$ in $S_0$.
$R_1[\frac{i}{3}]$ is the rank of the suffix $T[i+1,n)$ and $R_2[\frac{i}{3}]$ is the rank of suffix $T[i+2,n)$, which are both in the difference cover.
We write the tuple as $(i,c_0,c_1,r_1,r_2)$ where the indexes are interpreted relative to $i \bmod 3$.
Each suffix $i = 1 \bmod 3$ in $S_1$ stores $R_1[\frac{i-1}{3}]$, $R_2[\frac{i-1}{3}]$, and $T[i]$, and we write the tuples as $(i,c_1,r_1,r_2)$ where the indexes again are relative to $i \bmod 3$.
And lastly, each suffix $i = 2 \bmod 3$ in $S_2$ stores $R_1[\frac{i-2}{3}+1]$, $R_2[\frac{i-2}{3}]$, $T[i]$, and $T[i+1]$, because $R_1[\frac{i-2}{3}+1]$ is the rank of suffix $T[i+2,n)$.

In the Thrill algorithm code we construct the tuples by zipping $R_1$, $R_2$, and triple groups from $T$ together (line~25--29).
The Zip $Z'$ (line~28) delivers $(c_0,c_1,c_2,r_1,r_2)$ for each index $i = 0 \bmod 3$.
To construct the tuples in $S_i$ two adjacent tuples need to be used because $S_2$'s element are taken from the next tuple.
This can be done in Thrill using a $\Window$ operation of size $2$.
Thus to construct $S_0$, $S_1$, and $S_2$, we take $(c_0,c_1,c_2,r_1,r_2)$ for each index $i = 0 \bmod 3$ and $(\bar{c}_0,\bar{c}_1,\bar{c}_2,\bar{r}_1,\bar{r}_2)$ for the next index $i \bmod 3 + 3$, and output $(3i + 0,c_0,c_1,r_1,r_2)$ for $S_0$, $(3i + 1,c_0,c_1,r_1,c_2,r_2)$ for $S_1$, and $(3i + 2,c_2,r_2,\bar{c}_0,\bar{r}_1)$ for $S_2$, as described above (lines~30--32).
The three arrays are then sorted (lines~33--35) and merged, whereby the comparison functions compares two representatives character-wise until a rank is found.
The difference cover property guarantees that such a rank is found for every pair $S_i$, $S_j$ during the $\Merge$ (lines~36--44).

The difference cover algorithm DC3 generalizes to DC$X$ using a difference cover $D$ for any ground set size $X > 3$.
DC$X$ constructs a recursive subproblem of size $|D|/X$, has at most $\log_X |T|$ recursion levels and only one recursion branch.
At every level of the recursion, only work with sorting complexity is needed, and a straight-forward application of the Master theorem shows that the whole algorithms has the same complexity due to the small recursive subproblem.
In the RAM model and with integer alphabets one can use radix sort in each level and thus DC$X$ has linear running time.
For our distributed model, DC$X$ has the same complexity as sorting and merging.

Due to the subproblem size $|D|/X$ is it best to use the largest $X$ for a specific difference cover size.
For $|D| = 2$ this is $X = 3$, aka DC3.
For difference covers of size three, the largest $X = 7$ which yields DC7 with $D_7 = \{ 0, 1, 3 \}$.
And for difference covers of size four, the largest $X = 13$.
Weese~\cite{weese06entwurf} showed that DC7 is optimal regarding the number of I/Os in an external memory model assuming index types are four times the byte size of characters.
Due to these previous results we also implemented DC7 in Thrill.

Most of the previous discussion on DC3 can easily be extended to DC7: sort by seven characters instead of three, construct $T_R = T_0 \oplus T_1 \oplus T_3$ in case not all character tuples are unique, and have Step 1 deliver $R_0$, $R_1$, and $R_3$ containing the ranks of all suffixes $i \in D_7$.
We included the Thrill algorithm code for DC7 in Algorithms~\ref{alg:dc7-part1}--\ref{alg:dc7-cmp}.

The key to implementing DC7 is in the construction of the tuple contents of the seven arrays $S_0,\ldots,S_6$ from $R_0$, $R_1$, $R_3$, and characters from $T$.
Figure~\ref{fig:tuple-construction} shows a schematic to illustrate the underlying construction.
For each index $i$ there are three indexes $(i + k_0 \bmod 7)$, $(i + k_1 \bmod 7)$, and $(i + k_3 \bmod 7)$ in the difference cover $D_7$.
The offsets depend on $j = i \bmod 3$ for some index $i$, which classifies the suffix into $S_j$.
The tuples in the arrays must contain all characters up to (but excluding) the last known rank, since this is the information needed for the comparison function to perform character-wise comparisons up to the next known rank.
The components of the tuples in $S_0,\ldots,S_6$ visualized in Figure~\ref{fig:tuple-construction} are selected in Algorithm~\ref{alg:dc7-part2} from $Z$ via seven $\Map$ operations (lines~7--13).
They are then sorted by characters up to the next known rank (lines~14--20) and then merged using $\operatorname{CompareDC7}$ (Algorithm~\ref{alg:dc7-cmp}), which compares tuples character-wise up to the next known rank from all possible $S_i$/$S_j$ pairs.

In our Thrill implementation, $\operatorname{CompareDC7}$ is not rolled out as shown in the figure.
Instead a lookup tables is used to determine how many characters and which of the included ranks need to be compared.
Surprisingly, this more complex code was faster in our preliminary experiments, possibly due to the larger cost of decoding the instructions is the large unrolled comparison function.

\begin{figure}[t]
  \centering
  \begin{tikzpicture}
    \tikzset{
      marknode/.style={ every node/.append style={fill=blue!10} },
      finalnode/.style={ every node/.append style={fill=green!80!black!10} },
    }

    \matrix (m) [
      matrix of nodes,
      every node/.style={
        draw, anchor=base, align=center, text width=21pt, text height=8pt, text depth=2pt,
        name=\tikzmatrixname-\the\pgfmatrixcurrentrow-\the\pgfmatrixcurrentcolumn,
      },
      row sep=-\pgflinewidth, column sep=-\pgflinewidth,
      row 1 column 2/.style={marknode},
      row 1 column 3/.style={finalnode},
      row 2 column 1/.style={marknode},
      row 2 column 2/.style={finalnode},
      row 3 column 3/.style={finalnode},
      row 3 column 1/.style={marknode},
      ]
    {
      $c_0 $      & $c_1 / r_1$ & $r_2$       \\
      $c_1 / r_1$ & $r_2$       &             \\
      $c_2 / r_2$ & $\bar{c}_0$ & $\bar{r}_1$ \\
    };

    \node [left=1mm of m-1-1] {$S_0$};
    \node [left=1mm of m-2-1] {$S_1$};
    \node [left=1mm of m-3-1] {$S_2$};

    \node [below=1mm of m] {for DC3 with $D_3 = \{1,2\}$};

    \matrix (n) [
      right=15mm of m,
      matrix of nodes,
      every node/.style={
        draw, anchor=base, align=center, text width=21pt, text height=8pt, text depth=2pt,
        name=\tikzmatrixname-\the\pgfmatrixcurrentrow-\the\pgfmatrixcurrentcolumn,
      },
      row sep=-\pgflinewidth, column sep=-\pgflinewidth,
      row 1 column 1/.style={marknode},
      row 1 column 2/.style={marknode},
      row 1 column 4/.style={finalnode},
      row 2 column 7/.style={finalnode},
      row 2 column 1/.style={marknode},
      row 2 column 3/.style={marknode},
      row 3 column 6/.style={marknode},
      row 3 column 7/.style={finalnode},
      row 3 column 2/.style={marknode},
      row 4 column 5/.style={marknode},
      row 4 column 6/.style={finalnode},
      row 4 column 1/.style={marknode},
      row 5 column 4/.style={marknode},
      row 5 column 5/.style={marknode},
      row 5 column 7/.style={finalnode},
      row 6 column 3/.style={marknode},
      row 6 column 4/.style={marknode},
      row 6 column 6/.style={finalnode},
      row 7 column 2/.style={marknode},
      row 7 column 3/.style={marknode},
      row 7 column 5/.style={finalnode},
      ]
    {
      $c_0 / r_0$ & $c_1 / r_1$             & $c_2$                   & $r_3$                   &                         &                         &             \\
      $c_1 / r_1$ & $c_2$                   & $c_3 / r_3$             & $c_4$                   & $c_5$                   & $c_6$                   & $\bar{r}_0$ \\
      $c_2$       & $c_3 / r_3$             & $c_4$                   & $c_5$                   & $c_6$                   & $\bar{c}_0 / \bar{r}_0$ & $\bar{r}_1$ \\
      $c_3 / r_3$ & $c_4$                   & $c_5$                   & $c_6$                   & $\bar{c}_0 / \bar{r}_0$ & $\bar{r}_1$             &             \\
      $c_4$       & $c_5$                   & $c_6$                   & $\bar{c}_0 / \bar{r}_0$ & $\bar{c}_1 / \bar{r}_1$ & $\bar{c}_2$             & $\bar{r}_3$ \\
      $c_5$       & $c_6$                   & $\bar{c}_0 / \bar{r}_0$ & $\bar{c}_1 / \bar{r}_1$ & $\bar{c}_2$             & $\bar{r}_3$             &             \\
      $c_6$       & $\bar{c}_0 / \bar{r}_0$ & $\bar{c}_1 / \bar{r}_1$ & $\bar{c}_2$             & $\bar{r}_3$             &                         &             \\
    };

    \node [left=1mm of n-1-1] {$S_0$};
    \node [left=1mm of n-2-1] {$S_1$};
    \node [left=1mm of n-3-1] {$S_2$};
    \node [left=1mm of n-4-1] {$S_3$};
    \node [left=1mm of n-5-1] {$S_4$};
    \node [left=1mm of n-6-1] {$S_5$};
    \node [left=1mm of n-7-1] {$S_6$};

    \node [below=1mm of n] {for DC7 with $D_7 = \{0,1,3\}$};

  \end{tikzpicture}
  \caption{Construction of tuples in arrays $S_i$ to represent suffixes in DC3 and DC7.}\label{fig:tuple-construction}
\end{figure}

\begin{algorithm2e}
  \SetKwFunction{MakeTuples}{MakeTuples}
  \caption{DC7 Algorithm in Thrill (part one).\label{alg:dc7-part1}}
  \Function{DC7PartOne($T \in \DIA{\Sigma}$)}{
    $T_7 := T.\FlatWindow_7((i, \arr{c_0,c_1,\ldots,c_6}) \mapsto \MakeTuples(i,c_0,c_1,\ldots,c_6))$ \;
    \KwSty{with} \Function{\MakeTuples($i \in \mathbb{N}_0$, $c_0,c_1,\ldots,c_6 \in \Sigma$)}{
      \lIf(\Remi{Make tuples in difference cover.}){$i \in D_7$}{
        \KwSty{emit} $(i,c_0,c_1,\ldots,c_6)$
      }
    }
    $S := T_7.\Sort((i,c_0,c_1,\ldots,c_6) \text{ by } (c_0,c_1,\ldots,c_6))$ \Rem{Sort tuples lexicographically.}
    $I_S := S.\Map((i,c_0,c_1,\ldots,c_7) \mapsto i)$ \Rem{Extract sorted indices.}
    $N' := S.\FlatWindow_2((i, \arr{p_0,p_1}) \mapsto \operatorname{CmpTuple}(i,p_0,p_1))$ \Rem{Compare tuples.}
    \KwSty{with} \Function{CmpTuple($i \in \mathbb{N}_0$, $p_0 = (c_0,c_1,\ldots,c_6)$, $p_1 = (c'_0,c'_1,\ldots,c'_6))$)}{
      \lIf(\Remi{Emit one sentinel for index 0.}){$i = 0$}{
        \KwSty{emit} $0$
      }
      \lIf(\Remi{Emit 0 or 1 depending on}){$(c_0,c_1,\ldots,c_6) = (c'_0,c'_1,\ldots,c'_6)$}{
        \KwSty{emit} $0$
      }
      \lElse(\Remi{whether the previous tuple is equal.}){
       \KwSty{emit} $1$
      }
    }
    $N := N'.\PrefixSum()$ \Rem{Use prefix sum to calculate names.}
    $n_{\text{sub}} = \lceil 3 |T| / 7 \rceil$, \quad $n_{\text{mod0}} = \lceil |T| / 7 \rceil$ \Rem{Size of recursive problem and mod 0,}
    $n_{\text{mod1}} = \lceil |T| / 7 \rceil$, \quad $n_{\text{mod01}} = n_{\text{mod0}} + n_{\text{mod1}}$ \Rem{mod 1 and both parts of $T_R$.}
    \uIf(\Remi{If duplicate names exist, sort names back to}){$N.\Max() + 1 = n_{\text{sub}}$}{
      $T'_R := \Zip(\arr{ I_S, N }, (i, n) \mapsto (i,n)).\Sort((i,n) \text{ by } (i \bmod 7, i \text{ div } 7))$ \Rem{string order}
      $\SA_R := \operatorname{DC7}(T'_R.\Map( (i,n) \mapsto n ))$ \Rem{as $T_0 \oplus T_1 \oplus T_3$ and call suffix sorter.}
      $I'_R := \SA_R.\ZipWithIndex((i,r) \mapsto (i,r))$ \Rem{Invert resulting suffix array, but}
      $I_R := I'_R.\Sort((i,r) \text{ by } (\operatorname{InterleavedRank}(i), i))$ \Rem{interleaved $\ISA$ for better locality}
      \KwSty{with} \Function{InterleavedRank($i \in \mathbb{N}_0$)}{
        \Return (\KwSty{if} $i < n_{\text{mod0}}$ \KwSty{then} $i$ \KwSty{else if} $i < n_{\text{mod01}}$ \KwSty{then} $i - n_{\text{mod0}}$ \KwSty{else} $i - n_{\text{mod01}}$)\;
      }
      $R_0 := I_R.\Filter((i,r) \mapsto i < n_{\text{mod0}}).\Map((i,r) \mapsto r + 1)$ \Rem{after separating}
      $R_1 := I_R.\Filter((i,r) \mapsto i \geq n_{\text{mod0}} \text{ and } i < n_{\text{mod01}}).\Map((i,r) \mapsto r + 1)$ \Rem{$\ISA$ into}
      $R_3 := I_R.\Filter((i,r) \mapsto i \geq n_{\text{mod01}}).\Map((i,r) \mapsto r + 1)$ \Rem{$R_0$, $R_1$, and $R_3$.}
    }
    \Else(\Remi{Else, if all names/tuples are unique, then $I_S$ is already the suffix array.}){
      $R := I_S.\ZipWithIndex((i,r) \mapsto (i,r))$ \Rem{Invert it to get $\ISA$, but}
      $I_R := R.\Sort( (i,r) \text{ by } (i \text{ div } 7, i) )$ \Rem{interleave $\ISA$ for}
      $R_0 := I_R.\Filter((i,r) \mapsto i = 0 \bmod 7).\Map((i,r) \mapsto r + 1)$ \Rem{better locality}
      $R_1 := I_R.\Filter((i,r) \mapsto i = 1 \bmod 7).\Map((i,r) \mapsto r + 1)$ \Rem{after separating it}
      $R_3 := I_R.\Filter((i,r) \mapsto i = 3 \bmod 7).\Map((i,r) \mapsto r + 1)$ \Rem{into $R_0$, $R_1$, and $R_3$.}
    }
    \Return $\operatorname{DC7PartTwo}(T, R_0, R_1, R_3)$
  }
\end{algorithm2e}

\begin{algorithm2e}
  \SetKwFunction{MakeTuples}{MakeTuples}
  \SetKwFunction{CmpTuple}{CmpTuple}
  \caption{DC7 Algorithm in Thrill (part two).\label{alg:dc7-part2}}
  \Function{DC7PartTwo($T \in \DIA{\Sigma}$, $R_0, R_1, R_3 \in \DIA{\mathbb{N}_0}$)}{
    $\hat{T}_7 := T.\FlatWindow_7((i, \arr{c_0,c_1,\ldots,c_6}) \mapsto \MakeTuples(i,c_0,c_1,\ldots,c_6))$ \;
    \KwSty{with} \Function{\MakeTuples($i \in \mathbb{N}_0$, $c_0,c_1,\ldots,c_6 \in \Sigma$) \Remi{Prepare Zip with all}}{
      \lIf(\Remi{triples $i \notin D_7$.}){\KwSty{not} $i \notin D_7$}{
        \KwSty{emit} $(i,c_0,c_1,\ldots,c_6)$
      }
    }
    $Z' := \Zip(\arr{ \hat{T}_7,R_0,R_1,R_3 }, ((i,c_0,\ldots,c_6), r_0, r_1,r_3) \mapsto (c_0,\ldots,c_6,r_0,r_1,r_3)$\Rem{Pull}
    $Z := Z'.\Window_2( (i, \arr{ (z_1,z_2) }) \mapsto (i, z_1, z_2) )$ \Rem{chars and ranks using Zip from}

    $\begin{aligned}[t]
      S'_0 := Z.\Map( (i, (c_0,\ldots,c_6, & r_0,r_1,r_3),  (\bar{c}_0,\ldots,\bar{c}_6,\bar{r}_0,\bar{r}_1,\bar{r}_3)) \\
                                          & \mapsto (7 i + 0, c_0, r_0, c_1, r_1, c_2, r_3) )
      \end{aligned}$ \Rem{four arrays}
    $\begin{aligned}[t]
      S'_1 := Z.\Map( (i, (c_0,\ldots,c_6, & r_0,r_1,r_3),  (\bar{c}_0,\ldots,\bar{c}_6,\bar{r}_0,\bar{r}_1,\bar{r}_3)) \\
                                          & \mapsto (7 i + 1, c_1, r_1, c_2, c_3, r_3, c_4, c_5, c_6, \bar{r}_0)
      \end{aligned}$ \Rem{to make}
    $\begin{aligned}[t]
      S'_2 := Z.\Map( (i, (c_0,\ldots,c_6, & r_0,r_1,r_3),  (\bar{c}_0,\ldots,\bar{c}_6,\bar{r}_0,\bar{r}_1,\bar{r}_3)) \\
                                          & \mapsto (7 i + 2, c_2, c_3, r_3, c_4, c_5, c_6, \bar{c}_0, \bar{r}_0, \bar{r}_1)
      \end{aligned}$ \Rem{arrays of}
    $\begin{aligned}[t]
      S'_3 := Z.\Map( (i, (c_0,\ldots,c_6, & r_0,r_1,r_3),  (\bar{c}_0,\ldots,\bar{c}_6,\bar{r}_0,\bar{r}_1,\bar{r}_3)) \\
                                          & \mapsto (7 i + 3, c_3, r_3, c_4, c_5, c_6, \bar{c}_0, \bar{r}_0, \bar{r}_1)
      \end{aligned}$ \Rem{representatives}
    $\begin{aligned}[t]
      S'_4 := Z.\Map( (i, (c_0,\ldots,c_6, & r_0,r_1,r_3),  (\bar{c}_0,\ldots,\bar{c}_6,\bar{r}_0,\bar{r}_1,\bar{r}_3)) \\
                                          & \mapsto (7 i + 4, c_4, c_5, c_6, \bar{c}_0, \bar{r}_0, \bar{c}_1, \bar{r}_1, \bar{c}_2, \bar{r}_3)
      \end{aligned}$ \Rem{for each}
    $\begin{aligned}[t]
      S'_5 := Z.\Map( (i, (c_0,\ldots,c_6, & r_0,r_1,r_3),  (\bar{c}_0,\ldots,\bar{c}_6,\bar{r}_0,\bar{r}_1,\bar{r}_3)) \\
                                          & \mapsto (7 i + 5, c_5, c_6, \bar{c}_0, \bar{r}_0, \bar{c}_1, \bar{r}_1, \bar{c}_2, \bar{r}_3)
      \end{aligned}$ \Rem{suffix class.}
    $\begin{aligned}[t]
      S'_6 := Z.\Map( (i, (c_0,\ldots,c_6, & r_0,r_1,r_3),  (\bar{c}_0,\ldots,\bar{c}_6,\bar{r}_0,\bar{r}_1,\bar{r}_3)) \\
                                          & \mapsto (7 i + 6, c_6, \bar{c}_0, \bar{r}_0, \bar{c}_1, \bar{r}_1, \bar{c}_2, \bar{r}_3)
      \end{aligned}$ \;

    $S_0 := S'_0.\Sort( (i, c_0, r_0, c_1, r_1, c_2, r_3) \text{ by } (r_0) )$ \Rem{Sort representatives}
    $S_1 := S'_1.\Sort( (i, c_1, r_1, c_2, c_3, r_3, c_4, c_5, c_6, \bar{r}_0) \text{ by } (r_1) )$ \Rem{character-wise up to}
    $S_2 := S'_2.\Sort( (i, c_2, c_3, r_3, c_4, c_5, c_6, \bar{c}_0, \bar{r}_0, \bar{r}_1) \text{ by } (c_2,r_3) )$ \Rem{next rank, and merge}
    $S_3 := S'_3.\Sort( (i, c_3, r_3, c_4, c_5, c_6, \bar{c}_0, \bar{r}_0, \bar{r}_1) \text{ by } (r_3) )$ \Rem{sorted representatives}
    $S_4 := S'_4.\Sort( (i, c_4, c_5, c_6, \bar{c}_0, \bar{r}_0, \bar{c}_1, \bar{r}_1, \bar{c}_2, \bar{r}_3) \text{ by } (c_4,c_5,c_6,\bar{r}_0) )$ \Rem{to deliver the}
    $S_5 := S'_5.\Sort( (i, c_5, c_6, \bar{c}_0, \bar{r}_0, \bar{c}_1, \bar{r}_1, \bar{c}_2, \bar{r}_3) \text{ by } (c_5,c_6,\bar{r}_0) )$ \Rem{final suffix array.}
    $S_6 := S'_6.\Sort( (i, c_6, \bar{c}_0, \bar{r}_0, \bar{c}_1, \bar{r}_1, \bar{c}_2, \bar{r}_3) \text{ by } (c_6,\bar{r}_0) )$ \Rem{See Algorithm~\ref{alg:dc7-cmp}}

    \Return $\Merge(\arr{ S_0,S_1,\ldots,S_6 }, \operatorname{CompareDC7}).\Map((i,\ldots) \mapsto i)$ \Rem{for $\operatorname{CompareDC7}$.}
  }
\end{algorithm2e}

\begin{algorithm2e}
  \caption{Full Comparison Function in DC7.\label{alg:dc7-cmp}}
  \def\CmpLine[#1&#2&#3]{
    \parbox[b]{23ex}{#1}\quad$<$\quad\parbox[b]{23ex}{#2}\quad\parbox[b]{30ex}{if #3,}
  }
  \Function{CompareDC7($z_1,z_2$)}{
    \CmpLine[$(r_0)$                                     & $(r'_0)$                                                       & $z_1 \in S_0, z_2 \in S_0$] \;
    \CmpLine[$(r_0)$                                     & $(r'_1)$                                                       & $z_1 \in S_0, z_2 \in S_1$] \;
    \CmpLine[$(c_0,r_1)$                                 & $(c'_2,r'_3)$                                                  & $z_1 \in S_0, z_2 \in S_2$] \;
    \CmpLine[$(r_0)$                                     & $(r'_3)$                                                       & $z_1 \in S_0, z_2 \in S_3$] \;
    \CmpLine[$(c_0,c_1,c_2,r_3)$                         & $(c'_4,c'_5,c'_6,\bar{r}'_0)$                                  & $z_1 \in S_0, z_2 \in S_4$] \;
    \CmpLine[$(c_0,c_1,c_2,r_3)$                         & $(c'_5,c'_6,\bar{c}'_0,\bar{r}'_1)$                            & $z_1 \in S_0, z_2 \in S_5$] \;
    \CmpLine[$(c_0,r_1)$                                 & $(c'_6,\bar{r}'_0)$                                            & $z_1 \in S_0, z_2 \in S_6$] \;\vspace{4pt}
    \CmpLine[$(r_1)$                                     & $(r'_1)$                                                       & $z_1 \in S_1, z_2 \in S_1$] \;
    \CmpLine[$(c_1,c_2,c_3,c_4,c_5,c_6,\bar{r}_0)$       & $(c'_2,c'_3,c'_4,c'_5,c'_6,\bar{c}'_0,\bar{r}'_1)$             & $z_1 \in S_1, z_2 \in S_2$] \;
    \CmpLine[$(r_1)$                                     & $(r'_3)$                                                       & $z_1 \in S_1, z_2 \in S_3$] \;
    \CmpLine[$(c_1,c_2,c_3,c_4,c_5,c_6,\bar{r}_0)$       & $(c'_4,c'_5,c'_6,\bar{c}'_0,\bar{c}'_1,\bar{c}'_2,\bar{r}'_3)$ & $z_1 \in S_1, z_2 \in S_4$] \;
    \CmpLine[$(c_1,c_2,r_3)$                             & $(c'_5,c'_6,\bar{r}'_0)$                                       & $z_1 \in S_1, z_2 \in S_5$] \;
    \CmpLine[$(c_1,c_2,r_3)$                             & $(c'_6,\bar{c}'_0,\bar{r}'_1)$                                 & $z_1 \in S_1, z_2 \in S_6$] \;\vspace{4pt}
    \CmpLine[$(c_2,r_3)$                                 & $(c'_2,r'_3)$                                                  & $z_1 \in S_2, z_2 \in S_2$] \;
    \CmpLine[$(c_2,c_3,c_4,c_5,c_6,\bar{r}_0)$           & $(c_3,c_4,c_5,c_6,\bar{c}_0,\bar{r}'_1)$                       & $z_1 \in S_2, z_2 \in S_3$] \;
    \CmpLine[$(c_2,c_3,c_4,c_5,c_6,\bar{c}_0,\bar{r}_1)$ & $(c'_4,c'_5,c'_6,\bar{c}'_0,\bar{c}'_1,\bar{c}'_2,\bar{r}'_3)$ & $z_1 \in S_2, z_2 \in S_4$] \;
    \CmpLine[$(c_2,c_3,c_4,c_5,c_6,\bar{r}_0)$           & $(c'_5,c'_6,\bar{c}'_0,\bar{c}'_1,\bar{c}'_2,\bar{r}'_3)$      & $z_1 \in S_2, z_2 \in S_5$] \;
    \CmpLine[$(c_2,r_3)$                                 & $(c'_6,\bar{r}'_0)$                                            & $z_1 \in S_2, z_2 \in S_6$] \;\vspace{4pt}
    \CmpLine[$(r_3)$                                     & $(r'_3)$                                                       & $z_1 \in S_3, z_2 \in S_3$] \;
    \CmpLine[$(c_3,c_4,c_5,c_6,\bar{r}_0)$               & $(c'_4,c'_5,c'_6,\bar{c}'_0,\bar{r}'_1)$                       & $z_1 \in S_3, z_2 \in S_4$] \;
    \CmpLine[$(c_3,c_4,c_5,c_6,\bar{c}_0,\bar{r}_1)$     & $(c'_5,c'_6,\bar{c}_0,\bar{c}_1,\bar{c}_2,\bar{r}'_3)$         & $z_1 \in S_3, z_2 \in S_5$] \;
    \CmpLine[$(c_3,c_4,c_5,c_6,\bar{r}_0)$               & $(c'_6,\bar{c}'_0,\bar{c}'_1,\bar{c}'_2,\bar{r}'_3)$           & $z_1 \in S_3, z_2 \in S_6$] \;\vspace{4pt}
    \CmpLine[$(c_4,c_5,c_6,\bar{r}_0)$                   & $(c'_4,c'_5,c'_6,\bar{r}'_0)$                                  & $z_1 \in S_4, z_2 \in S_4$] \;
    \CmpLine[$(c_4,c_5,c_6,\bar{r}_0)$                   & $(c'_5,c'_6,\bar{c}'_0,\bar{r}'_1)$                            & $z_1 \in S_4, z_2 \in S_5$] \;
    \CmpLine[$(c_4,c_5,c_6,\bar{c}_0,\bar{r}_0)$         & $(c'_6,\bar{c}'_0,\bar{c}'_1,\bar{c}'_2,\bar{r}'_3)$           & $z_1 \in S_4, z_2 \in S_6$] \;\vspace{4pt}
    \CmpLine[$(c_5,c_6,\bar{r}_0)$                       & $(c'_5,c'_6,\bar{r}'_0)$                                       & $z_1 \in S_5, z_2 \in S_5$] \;
    \CmpLine[$(c_6,\bar{c}_0,\bar{r}_1)$                 & $(c'_6,\bar{c}'_0,\bar{r}'_1)$                                 & $z_1 \in S_5, z_2 \in S_6$] \;\vspace{4pt}
    \CmpLine[$(c_6,\bar{r}_0)$                           & $(c'_6,\bar{r}'_0)$                                            & $z_1 \in S_6, z_2 \in S_6$] \;\vspace{4pt}
    and symmetrically for $z_1 \in S_i, z_2 \in S_j$ if $i > j$\,, \;\vspace{4pt}
    with $z_1 = (i, c_0, r_0, c_1, r_1, c_2, r_3)$ if $z_1 \in S_0$, \;
    \qquad $z_2 = (i', c'_0, r'_0, c'_1, r'_1, c'_2, r'_3)$ if $z_2 \in S_0$, \;
    $z_1 = (i, c_1, r_1, c_2, c_3, r_3, c_4, c_5, c_6, \bar{r}_0)$ if $z_1 \in S_1$, \;
    \qquad $z_2 = (i', c'_1, r'_1, c'_2, c'_3, r'_3, c'_4, c'_5, c'_6, \bar{r}'_0)$ if $z_2 \in S_1$, \;
    $z_1 = (i, c_2, c_3, r_3, c_4, c_5, c_6, \bar{c}_0, \bar{r}_0, \bar{r}_1)$ if $z_1 \in S_2$, \;
    \qquad $z_2 = (i', c'_2, c'_3, r'_3, c'_4, c'_5, c'_6, \bar{c}'_0, \bar{r}'_0, \bar{r}'_1)$ if $z_2 \in S_2$, \;
    $z_1 = (i, c_3, r_3, c_4, c_5, c_6, \bar{c}_0, \bar{r}_0, \bar{r}_1)$ if $z_1 \in S_3$, \;
    \qquad $z_2 = (i', c'_3, r'_3, c'_4, c'_5, c'_6, \bar{c}'_0, \bar{r}'_0, \bar{r}'_1)$ if $z_2 \in S_3$,\;
    $z_1 = (i, c_4, c_5, c_6, \bar{c}_0, \bar{r}_0, \bar{c}_1, \bar{r}_1, \bar{c}_2, \bar{r}_3)$ if $z_1 \in S_4$, \;
    \qquad $z_2 = (i, c'_4, c'_5, c'_6, \bar{c}'_0, \bar{r}'_0, \bar{c}'_1, \bar{r}'_1, \bar{c}'_2, \bar{r}'_3)$ if $z_2 \in S_4$, \;
    $z_1 = (i, c_5, c_6, \bar{c}_0, \bar{r}_0, \bar{c}_1, \bar{r}_1, \bar{c}_2, \bar{r}_3)$ if $z_1 \in S_5$, \;
    \qquad $z_2 = (i', c'_5, c'_6, \bar{c'}_0, \bar{r}'_0, \bar{c}'_1, \bar{r}'_1, \bar{c}'_2, \bar{r}'_3)$ if $z_2 \in S_5$, \;
    $z_1 = (i, c_6, \bar{c}_0, \bar{r}_0, \bar{c}_1, \bar{r}_1, \bar{c}_2, \bar{r}_3)$ if $z_1 \in S_6$, \;
    \qquad $z_2 = (i', c'_6, \bar{c}'_0, \bar{r}'_0, \bar{c}'_1, \bar{r}'_1, \bar{c}'_2, \bar{r}'_3)$ if $z_2 \in S_6$. \;
  }
\end{algorithm2e}

\section{Conclusion}
We presented the implementation of five different suffix array construction algorithms in Thrill showing that the small set of algorithmic primitives provided by Thrill is sufficient to express the algorithms within the framework.

Our preliminary experimental results show that algorithms implemented in Thrill are competitive to hand-coded MPI implementations.
By using the Thrill framework we gain additional benefits like future improvements of the algorithmic primitives in Thrill, and possibly even fault tolerance.
Furthermore, Thrill already has automatic external memory support, hence our implementations are the first distributed external memory suffix array construction algorithms.

In a future version of this paper, we are going to add experimental results which detail the performance of our algorithms implemented in Thrill against their counterparts using MPI.

We will also show how compressed indexes like the FM-index can be efficiently constructed using the Thrill framework.
Additionally, we want to extend the existing algorithms with longest common prefix (LCP) array construction and the DC$X$ algorithms with discarding tuples~\cite{puglisi2005performance} similar to the technique we applied to the prefix doubling algorithms.

\bibliographystyle{plain}
\bibliography{local}

\end{document}